%% file: manuscript_sr_ADMM.tex
\documentclass[10pt,journal,compsoc]{IEEEtran}
\ifCLASSOPTIONcompsoc
  \usepackage[nocompress]{cite}
\else
  \usepackage{cite}
\fi
\ifCLASSINFOpdf
\else
\fi

\usepackage{amssymb}
\usepackage{mathtools}
\usepackage[english]{babel}
\usepackage{amsmath}
\usepackage{amsthm}  
\usepackage{url}

\usepackage{algorithm}
\usepackage{algpseudocode}

\newcounter{MYtempeqncnt}

\theoremstyle{plain}
\newtheorem{thm}{Theorem}[section]

\newtheorem{assump}[thm]{Assumption}

\theoremstyle{definition}

\theoremstyle{remark}
\newtheorem{remark}{Remark}[section]

\include{macros}

\DeclareMathOperator{\diag}{diag}
\DeclareMathOperator{\sign}{sign}

\newcommand{\Prob}[1]{\Pr\left(#1\right)}
\newcommand{\abs}[1]{\left|#1\right|}
\newcommand{\twonorm}[1]{\normof{#1}{}}                  

\newcommand{\rev}[1]{{\color[rgb]{0,0,0}{{#1}}}}

\begin{document}
%
\title{A Unified Model for Differential Expression Analysis of RNA-seq Data via L1-Penalized Linear Regression}
%
%
%
%

\author{Kefei Liu,~
        Jieping Ye,~\IEEEmembership{Senior Member,~IEEE,}
        Yang Yang,~
        Li Shen,~\IEEEmembership{Member,~IEEE,}
        and~Hui Jiang
\IEEEcompsocitemizethanks{\IEEEcompsocthanksitem Kefei Liu and Li Shen are with the Department of Radiology and Imaging Sciences, Indiana University School of Medicine, Indianapolis, IN 46202.\protect\\
E-mail: \{kefliu,shenli\}@iu.edu
\IEEEcompsocthanksitem Jieping Ye is with the Department of Computational Medicine and Bioinformatics, University of Michigan, MI 48109.\protect\\
E-mail: jpye@umich.edu
\IEEEcompsocthanksitem Yang Yang is with the School of Computer Science and Engineering, Beihang University, Beijing 100191, China.\protect\\
E-mail: yangyangfuture@buaa.edu.cn
\IEEEcompsocthanksitem Hui Jiang is with the Department of Biostatistics, University of Michigan, MI 48109.
E-mail: jianghui@umich.edu}}

%
%

\markboth{IEEE/ACM Transactions on Computational Biology and Bioinformatics}
{Shell \MakeLowercase{\textit{et al.}}: Bare Demo of IEEEtran.cls for Computer Society Journals}
%



\IEEEtitleabstractindextext{%
\begin{abstract}
The RNA-sequencing (RNA-seq) is becoming increasingly popular for quantifying gene expression levels. Since the RNA-seq measurements are relative in nature, between-sample normalization of counts is an essential step in differential expression (DE) analysis. The normalization of existing DE detection algorithms is ad hoc and performed once for all prior to DE detection, which may be suboptimal since ideally normalization should be based on non-DE genes only and thus coupled with DE detection. We propose a unified statistical model for joint normalization and DE detection of log-transformed RNA-seq data. Sample-specific normalization factors are modeled as unknown parameters in the gene-wise linear models and jointly estimated with the regression coefficients. By imposing sparsity-inducing L1 penalty (or mixed L1/L2-norm for multiple treatment conditions) on the regression coefficients, we formulate the problem as a penalized least-squares regression problem and apply the augmented lagrangian method to solve it. Simulation studies show that the proposed model and algorithms outperform existing methods in terms of detection power and false-positive rate when more than half of the genes are differentially expressed and/or when the up- and down-regulated genes among DE genes are unbalanced in amount.
\end{abstract}

}

\maketitle

\IEEEdisplaynontitleabstractindextext

%
\IEEEpeerreviewmaketitle

\IEEEraisesectionheading{\section{Introduction}\label{sec:intro}}
Ultra high-throughput sequencing of transcriptomes (RNA-seq) is a widely used method for quantifying gene expression levels due to its low cost, high accuracy and wide dynamic range for detection~\cite{Mortazavi2008}. As of today, modern ultra high-throughput sequencing platforms can generate hundreds of millions of sequencing reads from each biological sample in a single day. RNA-seq also facilitates the detection of novel transcripts~\cite{Trapnell2010} and the quantification of transcripts on isoform level~\cite{Jiang2009,Salzman2011}. For these reasons, RNA-seq has become the method of choice for assaying transcriptomes~\cite{Wang2009}.

One major limitation of RNA-seq is that it only provides relative measurements of transcript abundances due to difference in library size (i.e., sequencing depth) between samples~\cite{Dillies2013}. Normalization of RNA-seq read counts is required in gene differential expression analysis to correct for such variation between samples. A popular form of between-sample normalization is achieved by scaling raw read counts in each sample by a sample-specific factor related to library size~\cite{Dillies2013,Rapaport2013}. This include CPM/RPM (counts/reads per million)~\cite{Robinson2010a}, quantile normalization~\cite{Bolstad2003,Smyth2005}, upper-quartile normalization~\cite{Bullard2010}, trimmed mean of M values~\cite{Robinson2010a} and DESeq normalization~\cite{Anders2010}. Also, commonly-used gene expression measures, e.g., TPM (transcript per million)~\cite{Li2010a}, and RPKM/FPKM (reads/fragments per kilobase of exon per million mapped reads)~\cite{Mortazavi2008},\rev{\cite{Trapnell2010}}, also correct for difference in gene length within a sample~\cite{Oshlack2009} (the so-called within-sample normalization). 
In particular, the CPM/RPM (counts/reads per million)~\cite{Robinson2010a}, TPM (transcript per million)~\cite{Li2010a}, and RPKM/FPKM (reads/fragments per kilobase of exon per million mapped reads)~\cite{Mortazavi2008,Trapnell2010} for the $i$-th gene from the $j$-th sample are respectively defined as
\begin{equation}
\label{units}
\begin{array}{lll}
{\rm cpm}_{ij}&=& 10^6\frac{c_{ij}}{N_j}\\
{\rm fpkm}_{ij}&=& 10^9\frac{c_{ij}}{l_i \cdot N_j}\\
{\rm tpm}_{ij}&=& 10^6 \frac{c_{ij}/l_i}{\displaystyle{\sum_{i}} c_{ij}/l_i}
\end{array}
\end{equation}
where $c_{ij}$ is the observed read count for gene $i$ from the $j$-th sample, $N_j = \displaystyle{\sum_{i}} c_{ij}$ is the sequencing depth in the $j$-th sample, and $l_i$ be the length of gene $i$. 
In this work we focus on between-sample normalization.


In traditional count-based RNA-seq analysis methods, the read counts for each gene are assumed to follow a Poisson~\cite{Marioni2008} or negative binomial (NB) distribution. 
One issue with the count-based RNA-seq analysis methods is that their procedures are complicated and contain many ad hoc heuristics. Moreover, the Poisson or NB distributions of counts are mathematically less tractable than the normal distribution~\cite{limma-voom,limma}. This makes count-based methods difficult to generalize to new data. Moreover, commonly-used statistical methods for microarray data analysis, e.g., quality weighting of RNA samples, addition of random noise to generate technical replicates, and gene set test~\cite{limma-voom} have been designed for normally distributed data and it is unclear whether we can adapt them to count data. \rev{Also the presence of outliers is an issue that is not well addressed (addressed in a very ad hoc manner) by existing methods.} To handle that, the authors of~\cite{limma-voom} take the logarithm of the raw count of reads and apply normal distribution-based statistical methods to analyze them. Note that by logarithmic transformation, the dynamic range of the RNA-seq counts is compressed such that the outlier counts are largely transformed into ``normal" data. As a result, sophisticated way to detect and discard outliers\rev{~\cite{DESeq2,Jiang2015,edgeR-robust}} is not required.

In this paper, like in~\cite{limma-voom,limma} we work with log-transformed gene expression values and propose a unified statistical model for differential gene expression. Different from~\cite{limma,limma-voom}, we model sample-specific scaling factors for between-sample normalization as unknown parameters and incorporate them into the gene-wise linear models. By imposing the sparsity-inducing penalty ($\ell_1$-norm for single treatment factor and mixed $\ell_1$/$\ell_2$-norm for multiple treatment factors) on the regression coefficients and carefully choosing the penalty parameter, the model is able to achieve joint accurate detection of DE genes and between-sample normalization. 
To fit the model, we first eliminate sample-specific parameters using optimization argumentation to formulate the problem as a penalized linear regression problem, and then solve it with the alternating direction method of multipliers algorithm (ADMM), which is known for its fast convergence to modest accuracy~\cite{Boyd2011}. Regarding the choice of penalty parameter, we theoretically derive the smallest penalty parameter $\alpha_{\rm max}$ that leads to all-zero solution, and thereby find a proper penalty parameter within $[0,\alpha_{\rm max}]$. Simulation studies show that \rev{the proposed methods perform better in terms of detection power and false-positive rate than existing methods when more than 50\% of the genes are differentially expressed and/or the fold change distribution is asymmetric \footnote[1]{Here by ``asymmetric" we mean the up- and down-regulated DE genes are unbalanced in number \rev{[(or amount)]}.}. Moreover, it is robust against deviations of the distribution of RNA-seq count data and suffers almost no performance degradation even when the data is generated according to the probabilistic assumptions of the previous methods.} 


Note that our work is preceded by \cite{Jiang2016} which address the differential expression problem in a similar way. \rev{The difference is that the model of \cite{Jiang2016} considers only categorical or qualitative predictor/explanatory variables (treatment conditions). For example, label ``0" is assigned to samples from the control group and label ``1" to samples from the treatment group.} While in our model, the precitor/explanatory variables can take arbitrary numeric values, and is thus a generalization of \cite{Jiang2016} from discrete to continuous predictor-variable model case. Note that the algorithm in \cite{Jiang2016} does not apply to the current numeric variable model at hand, because (i) applicability: it requires that multiple samples are present in each group but in the continuous-predictor model the concept of ``group" no longer exists, or more precisely, each group is formed by only one sample; (ii) algorithmic complexity: it requires an $p$-dimensional exhaustive search, where $p$ is the number of treatment conditions. When $p>1$ (see Section \ref{sec:multiple_regression_model}), the algorithm is computationally very expensive.




The remainder of the paper is organized as follows. In Section~\ref{sec:model}, we formulate the problem in the context of a single treatment factor. In Section~\ref{sec:alg}, we formulate the problem as a penalized simple regression problem and derive efficient ADMM algorithm to solve it, together with the estimation of noise variance and penalty parameter. In Section \ref{sec:multiple_regression_model}, we extend the simple regression model to multiple linear regression model. Comparison with existing methods is presented in Section~\ref{sec:experiments}, followed by discussions in Section~\ref{sec:discussion}.


\section{Data model and Problem Formulation}
\label{sec:model}

\rev{Throughout the paper, the subscript is used to index the vectors for rows of a matrix. For example, the $i$-th row vector of a matrix $\mat{A}$ is denoted as $\mat{a}_i$.}


\subsection{Data model}\label{subsec:model}
Suppose there are a total of $m$ genes measured in $n$ samples. Let $y_{ij}$, $i = 1,2,\dots,m$, $j=1,2,\dots,n$, be the log-transformed gene expression measurements (a small positive number is usually added before taking logarithm) for the $i$-th gene from the $j$-th sample. The following statistical model is assumed
\begin{equation}\label{equ:simple_regression_model}
    y_{ij} \sim \mathcal{N}\left(\beta_{0i}+\beta_i x_j +d_j,\sigma_i^2\right)
\end{equation}
where $\beta_{0i}$ is the $y$-intercept for gene $i$, $x_j$, $j=1,2,\dots,n$, is the predictor variable that represents the treatment condition (e.g., drug dosage) for sample $j$, $\beta_i$ is the slope or regression coefficient representing log-fold-change of expression levels of gene $i$ with unit change of $x_j$, $d_j$ is the scaling factor (e.g., $\log(\mbox{sequencing depth})$ or $\log(\mbox{library size})$) for sample $j$ for between-sample normalization~\cite{Dillies2013}, and $\sigma_i$ is the standard deviation of log-transformed expression levels of gene $i$.

In \eqref{equ:simple_regression_model}, we consider a single treatment condition. Extension to models with multiple treatment conditions will be discussed in Section~\ref{sec:multiple_regression_model}.






Our main interest is to detect differentially expressed (DE) genes, i.e., whether $\beta_i$ is equal to zero. If $\beta_i\neq 0$ gene $i$ is differentially expressed across the $n$ samples; otherwise it is not. 

\begin{remark}
Since $\beta_{0i}$ and $d_j$ in \eqref{equ:simple_regression_model} respectively model gene-specific factor (e.g., gene length) and sample-specific factor, model \eqref{equ:simple_regression_model} able to work with any log-transformed gene expression measures in the form of
\begin{equation}\label{equ:units_form}
    y_{ij} = \log \frac{c_{ij}}{l_i \cdot q_j},
\end{equation}
where $c_{ij}$ is the raw counts, $l_i$ is the length of gene $i$ and $q_j$ is the normalization factor of the $j$-th sample, since $l_i$ and $q_j$ can be absorbed into $\beta_{0i}$ and $d_i$, respectively. Note that gene expression measures of form $c_{ij} / (l_i \cdot q_j)$ include the raw counts (with $l_i=q_j=1$), measures based on between-sample normalization only ($l_i=1$)~\cite{Dillies2013}, and FPKM and TPM which are shown in \eqref{units} and involve both between- and within-sample normalization.
\end{remark}

\subsection{Penalized likelihood}
Since the gene expression measurements $y_{ij}$ are independent across the genes and samples, the likelihood is given by
\begin{equation}\label{equ:likelihood}
\Prob{\mat{y} | \mat{\beta}_0, \mat{\beta}, \mat{d}} = \displaystyle{\prod_{i=1}^m} \displaystyle{\prod_{j=1}^n} \frac{1}{\sqrt{2\pi \sigma_i^2}} \exp\left\{ -\frac{\left(y_{ij}-\beta_{0i}-\beta_i x_j-d_j\right)^2}{2\sigma_i^2} \right\},
\end{equation}
where
\begin{equation*}
    \mat{\beta} = \left(
                     \begin{array}{c}
                       \beta_1 \\
                       \beta_2 \\
                       \vdots \\
                       \beta_m \\
                     \end{array}
                   \right).
\end{equation*}


Assume that $ \{\sigma_i^2\}_{i=1}^m$ are known, maximization of \eqref{equ:likelihood} is equivalent to minimizing the negative log-likelihood:
\begin{equation}\label{equ:negative_log-likelihood}
l\left(\mat{\beta}_0, \mat{\beta}, \mat{d}; \mat{y}\right) = \displaystyle{\sum_{i=1}^m} \displaystyle{\sum_{j=1}^n} \frac{1}{2\sigma_i^2} \left(y_{ij}-\beta_{0i}-\beta_i x_j-d_j\right)^2,
\end{equation}
where we have ignored the irrelevant constant.

In practice, we solve for $ \{\sigma_i^2\}_{i=1}^m$ using an ad hoc approach, which will be described in Section~\ref{subsec:estimate_noise_variance}.

We introduce a $\ell_1$-penalty on the $\beta_i$'s:
\begin{equation}\label{equ:penalty_function}
    p(\beta) = \alpha \normof{\mat{\beta}}{1} \coloneqq \alpha \displaystyle{\sum_{i=1}^m} \, \abs{\beta_i}.
\end{equation}
It is well known that the $\ell_1$-norm penalty favors sparse solutions (forces some coefficients to be exactly zero)~\cite{Tibshirani1996}. This is reasonable since in practice many genes are not differentially expressed.

The objective function to be minimized is
\begin{equation}\label{equ:objective_function}
f\left(\mat{\beta}_0, \mat{\beta}, \mat{d}\right) = \displaystyle{\sum_{i=1}^m} \displaystyle{\sum_{j=1}^n} \frac{1}{2\sigma_i^2} \left(y_{ij}-\beta_{0i}-x_j \beta_i-d_j\right)^2 + \alpha \displaystyle{\sum_{i=1}^m} \, \abs{\beta_i}.
\end{equation}

\section{Algorithm Development}
\label{sec:alg}
\begin{figure*}[!b]
\normalsize
\vspace*{4pt}
\hrulefill
\setcounter{MYtempeqncnt}{\value{equation}}
\setcounter{equation}{27}
\begin{equation}\label{equ:objective_function_augmented_Lagrangian}
L_\rho\left(\mat{\beta}, \delta_0, \lambda\right) = \displaystyle{\sum_{i=1}^m} \frac{1}{2\sigma_i^2}
 \displaystyle{\sum_{j=1}^n} \left(\tilde{y}_{ij} - x_j \beta_i + x_j \delta_0 \right)^2 + \alpha \displaystyle{\sum_{i=1}^m} \, \abs{\beta_i}
 + \lambda \left(\frac{1}{\displaystyle{\sum_{i=1}^m} \frac{1}{\sigma_i^2}} \displaystyle{\sum_{i=1}^m} \frac{1}{\sigma_i^2} \beta_i - \delta_0\right)
+ \frac{\rho}{2} \left(\frac{1}{\displaystyle{\sum_{i=1}^m} \frac{1}{\sigma_i^2}} \displaystyle{\sum_{i=1}^m} \frac{1}{\sigma_i^2} \beta_i - \delta_0\right)^2.
\end{equation}
\setcounter{equation}{\value{MYtempeqncnt}}
\end{figure*}

\begin{figure*}[!b]
\normalsize
\vspace*{4pt}
\hrulefill
\setcounter{MYtempeqncnt}{\value{equation}}
\setcounter{equation}{29}
\begin{equation}\label{equ:beta_i_esti}
\beta_i = \frac{\sigma_i^2\left(\displaystyle{\sum_{\ell=1}^m} \sigma_\ell^{-2}\right)^2}{\sigma_i^2\left(\displaystyle{\sum_{\ell=1}^m} \sigma_\ell^{-2}\right)^2+\rho} T_{\sigma_i^2\alpha} \left[ \left(\sum_{j=1}^n x_j \tilde{y}_{ij} + \delta_0 \right) - \frac{\rho}{\displaystyle{\sum_{\ell=1}^m} \frac{1}{\sigma_\ell^2}} \left(\frac{1}{\displaystyle{\sum_{\ell=1}^m} \frac{1}{\sigma_\ell^2}} \displaystyle{\sum_{\ell\neq i}} \frac{1}{\sigma_\ell^2} \beta_\ell - \delta_0 + \frac{\lambda}{\rho}\right) \right],
\end{equation}
\setcounter{equation}{\value{MYtempeqncnt}}
\end{figure*}

\subsection{Formulation of \eqref{equ:objective_function} as Penalized Simple Linear Regression Model}
\label{subsec:simple_regression_model}

\rev{It can be proved that the optimization problem in \eqref{equ:objective_function} is jointly convex in $\left(\mat{\beta}_0, \mat{\beta}, \mat{d}\right)$. Therefore, the minimizer of \eqref{equ:objective_function} is the stationary point.}

The derivative of $f\left(\mat{\beta}_0, \mat{\beta}, \mat{d}\right)$ with respect to $d_j$, $j=1,2,\dots,n$, is
\begin{equation}\label{equ:derivative_of_f_with_respect_to_d_j}
    \frac{\partial f}{\partial d_j} = \displaystyle{\sum_{i=1}^m} -\frac{1}{\sigma_i^2} \left(y_{ij}-\beta_{0i}-x_j \beta_i-d_j\right) = 0.
\end{equation}

Setting \eqref{equ:derivative_of_f_with_respect_to_d_j} to zero gives
\begin{equation}\label{}
    d_j = \frac{1}{\displaystyle{\sum_{i=1}^m} \frac{1}{\sigma_i^2}} \displaystyle{\sum_{i=1}^m} \frac{1}{\sigma_i^2} \left(y_{ij}-\beta_{0i}-x_j \beta_i\right).
\end{equation}

Model \eqref{equ:simple_regression_model} is non-identifiable because we can simply add any constant to all the $d_j$'s and subtract the same constant from all the $\beta_{0i}$'s, while having the same fit. To resolve this issue, we fix $d_1 = 0$. Therefore
\begin{equation}\label{equ:d_j}
    d_j = d_j -d_1 = \left(\bar{y}_{\cdot j}^{(w)}-\bar{y}_{\cdot 1}^{(w)}\right) - \left(x_j-x_1\right) \bar{\beta}^{(w)},
\end{equation}
where
\begin{equation}\label{equ:weighted_average_y_over_i}
    \bar{y}_{\cdot j}^{(w)} \coloneqq \frac{1}{\displaystyle{\sum_{i=1}^m} \frac{1}{\sigma_i^2}} \displaystyle{\sum_{i=1}^m} \frac{1}{\sigma_i^2} y_{ij}, \mbox{ for } j=1,2,\dots,n,
\end{equation}
\begin{equation}\label{equ:weighted_average_beta_over_i}
    \bar{\beta}^{(w)} \coloneqq \frac{1}{\displaystyle{\sum_{i=1}^m} \frac{1}{\sigma_i^2}} \displaystyle{\sum_{i=1}^m} \frac{1}{\sigma_i^2} \beta_i.
\end{equation}

On the other hand, from
\begin{equation}\label{}
    \frac{\partial f}{\partial \beta_{0i}} = -\frac{1}{\sigma_i^2} \displaystyle{\sum_{j=1}^n} \left(y_{ij}-\beta_{0i}-x_j \beta_i-d_j\right) = 0,
\end{equation}
we have
\begin{equation}\label{equ:mu_i}
    \beta_{0i} = \frac{1}{n} \displaystyle{\sum_{j=1}^n} \left(y_{ij}-x_j \beta_i-d_j\right) = \bar{y}_{i\cdot} - \bar{x} \beta_i - \frac{1}{n} \displaystyle{\sum_{j=1}^n} d_j.
\end{equation}
where 
\begin{equation}\label{equ:average_y_over_j}
    \bar{y}_{i\cdot} \coloneqq \frac{1}{n} \displaystyle{\sum_{j=1}^n} y_{ij}, \mbox{ for } i=1,2,\dots,m.
\end{equation}
\begin{equation}\label{equ:def_of_xbar}
    \bar{x} \coloneqq \frac{1}{n} \displaystyle{\sum_{j=1}^n} x_j.
\end{equation}

From \eqref{equ:d_j} we have
\begin{equation}\label{equ:d_j_sum_over_j}
    \frac{1}{n} \displaystyle{\sum_{j=1}^n} d_j = \left(\bar{y}^{(w)} - \bar{y}_{\cdot 1}^{(w)}\right) - \left(\bar{x}-x_1\right) \bar{\beta}^{(w)},
\end{equation}
where
\begin{equation}\label{equ:weighted_average_y_over_ij}
    \bar{y}^{(w)} \coloneqq \frac{1}{\displaystyle{\sum_{i=1}^m} \frac{1}{\sigma_i^2}} \displaystyle{\sum_{i=1}^m} \frac{1}{\sigma_i^2} \cdot \frac{1}{n} \displaystyle{\sum_{j=1}^n} y_{ij}.
\end{equation}

Substituting \eqref{equ:d_j_sum_over_j} into \eqref{equ:mu_i} yields
\begin{equation}\label{equ:mu_i_no_d_j}
    \beta_{0i} = \bar{y}_{i\cdot} + \bar{y}_{\cdot 1}^{(w)} - \bar{y}^{(w)} + \left(\bar{x}-x_1\right) \bar{\beta}^{(w)} - \bar{x} \beta_i.
\end{equation}

Without loss of generality, we make the following assumptions:
\begin{assump}\label{equ:assum}
\begin{equation}
  \displaystyle{\sum_{j=1}^n} x_j = n \bar{x} = 0, \quad \displaystyle{\sum_{j=1}^n} x_j^2 = 1.
\end{equation}
\end{assump}
This assumption is reasonable since in the model \eqref{equ:simple_regression_model} the center and scaling factor of $x_j$'s can be absorbed into $\beta_{0i}$ and $\beta_i$, respectively.

Then \eqref{equ:mu_i_no_d_j} simplifies to
\begin{equation}\label{equ:mu_i_no_d_j_simplfied}
    \beta_{0i} = \bar{y}_{i\cdot} + \bar{y}_{\cdot 1}^{(w)} - \bar{y}^{(w)} - x_1 \bar{\beta}^{(w)}.
\end{equation}

The sum of \eqref{equ:d_j} and \eqref{equ:mu_i_no_d_j_simplfied} yields
\begin{equation}\label{equ:mu_i_plus_d_j}
    \beta_{0i} + d_j = \bar{y}_{i\cdot} + \bar{y}_{\cdot j}^{(w)} - \bar{y}^{(w)} - x_j \bar{\beta}^{(w)}.
\end{equation}

Substituting \eqref{equ:mu_i_plus_d_j} into \eqref{equ:objective_function}, the latter simplifies to
\begin{equation}\label{equ:objective_function_simplfied}
f\left(\mat{\beta}\right) = \displaystyle{\sum_{i=1}^m} \frac{1}{2\sigma_i^2}
 \displaystyle{\sum_{j=1}^n} \left(\tilde{y}_{ij} - x_j \beta_i + x_j \bar{\beta}^{(w)}\right)^2 + \alpha \displaystyle{\sum_{i=1}^m} \, \abs{\beta_i},
\end{equation}
where
\begin{equation}\label{expression:tilde_y_ij}
    \tilde{y}_{ij} \coloneqq y_{ij} - \bar{y}_{i\cdot} - \bar{y}_{\cdot j}^{(w)} + \bar{y}^{(w)}.
\end{equation}

It can be shown by straightforward calculation that $\{\tilde{y}_{ij}\}$ satisfies
\begin{equation}\label{equ:y_tilde_sum_over_i}
    \displaystyle{\sum_{i=1}^m} \frac{1}{\sigma_i^2} \tilde{y}_{ij} = 0.
\end{equation}
\begin{equation}\label{equ:y_tilde_sum_over_j}
    \displaystyle{\sum_{j=1}^n} \tilde{y}_{ij} = 0.
\end{equation}

\subsection{Model Fitting by ADMM}
\label{subsec:ADMM}


We propose to use the alternating direction method of multipliers (ADMM)~\cite{Boyd2011} to solve \eqref{equ:objective_function_simplfied}. 
Although ADMM can be very slow to converge to high accuracy, it is often the case that ADMM converges to modest accuracy very fast (within a few tens of iterations)~\cite{Boyd2011}.

To apply the ADMM, the problem \eqref{equ:objective_function_simplfied} is reformulated as
\begin{subequations}\label{equ:objective_function_simplfied_ADMM}
\begin{equation}
f\left(\mat{\beta}\right) = \displaystyle{\sum_{i=1}^m} \frac{1}{2\sigma_i^2}
 \displaystyle{\sum_{j=1}^n} \left(\tilde{y}_{ij} - x_j \beta_i + x_j \delta_0 \right)^2 + \alpha \displaystyle{\sum_{i=1}^m} \, \abs{\beta_i},
\end{equation}
subject to
\begin{eqnarray}
  \frac{1}{\displaystyle{\sum_{i=1}^m} \frac{1}{\sigma_i^2}} \displaystyle{\sum_{i=1}^m} \frac{1}{\sigma_i^2} \beta_i &=& \delta_0.
\end{eqnarray}
\end{subequations}

The augmented Lagrangian of \eqref{equ:objective_function_simplfied_ADMM} is \eqref{equ:objective_function_augmented_Lagrangian} \rev{at the bottom of the page.} \addtocounter{equation}{1}

Step 1: Update $\beta_i$, $i=1,2,\dots,m$:


The derivative of \eqref{equ:objective_function_augmented_Lagrangian} with respect to $\beta_i$ is
\begin{equation}\label{equ:derivatives_of_Li_with_respect_to_beta_i}
\begin{split}
\frac{\partial L_\rho}{\partial \beta_i} = &\frac{1}{\sigma_i^2} \displaystyle{\sum_{j=1}^n} - x_j \left(\tilde{y}_{ij} - x_j \beta_i + x_j \delta_0\right) + \alpha \partial \abs{\beta_i} \\
&+ \frac{1}{\displaystyle{\sum_{\ell=1}^m} \frac{1}{\sigma_\ell^2}} \frac{1}{\sigma_i^2} \lambda + \rho \frac{1}{\displaystyle{\sum_{\ell=1}^m} \frac{1}{\sigma_\ell^2}} \frac{1}{\sigma_i^2} \left(\frac{1}{\displaystyle{\sum_{\ell=1}^m} \frac{1}{\sigma_\ell^2}} \displaystyle{\sum_{\ell=1}^m} \frac{1}{\sigma_\ell^2} \beta_\ell - \delta_0\right),
\end{split}
\end{equation}
where $\partial \abs{\beta_i}$ is the subgradient of $\abs{\beta_i}$ with respect to $\beta_i$ and is defined as
\begin{equation*}
    \partial \abs{\beta_i} =
\begin{cases}
1, & \beta_i>0 \\
-1, & \beta_i<0 \\
[-1,1], & \beta_i=0
\end{cases}
\end{equation*}

Setting \eqref{equ:derivatives_of_Li_with_respect_to_beta_i} equal to zero gives \eqref{equ:beta_i_esti} \rev{at the bottom of the page,} \addtocounter{equation}{1}
where $T$ is the soft-thresholding operator:
\begin{equation*}
    T_{\sigma_i^2\alpha} \left[ x \right] \coloneqq \sign(x) \left( \abs{x}-\sigma_i^2\alpha \right)_{+} =
\begin{cases}
x-\sigma_i^2\alpha, & x>\sigma_i^2\alpha \\
x+\sigma_i^2\alpha, & x<-\sigma_i^2\alpha \\
0, & -\sigma_i^2\alpha \leq x \leq \sigma_i^2\alpha
\end{cases}
\end{equation*}

Step 2: Update $\delta_0$:

The derivative of \eqref{equ:objective_function_augmented_Lagrangian} with respect to $\delta_0$ is
\begin{equation}\label{equ:derivatives_of_Lrho_with_respect_to_delta_0}
\begin{split}
L_\rho\left(\mat{\beta}, \delta_0, \lambda\right) = &\displaystyle{\sum_{i=1}^m} \frac{1}{\sigma_i^2} \displaystyle{\sum_{j=1}^n} x_j \left(\tilde{y}_{ij} - x_j \beta_i + x_j \delta_0\right) \\
&- \lambda + \rho \left( \delta_0 - \frac{1}{\displaystyle{\sum_{i=1}^m} \frac{1}{\sigma_i^2}} \displaystyle{\sum_{i=1}^m} \frac{1}{\sigma_i^2} \beta_i \right).
\end{split}
\end{equation}

Setting \eqref{equ:derivatives_of_Lrho_with_respect_to_delta_0} equal to zero gives
\begin{equation}\label{equ:delta_0}
\begin{split}
\delta_0 &= \frac{1}{\displaystyle{\sum_{i=1}^m} \frac{1}{\sigma_i^2}+\rho} \left(\lambda - \displaystyle{\sum_{i=1}^m} \frac{1}{\sigma_i^2} \sum_{j=1}^n x_j \tilde{y}_{ij} \right) + \frac{1}{\displaystyle{\sum_{i=1}^m} \frac{1}{\sigma_i^2}} \displaystyle{\sum_{i=1}^m} \frac{1}{\sigma_i^2} \beta_i \\
&= \frac{1}{\displaystyle{\sum_{i=1}^m} \frac{1}{\sigma_i^2}+\rho} \lambda + \frac{1}{\displaystyle{\sum_{i=1}^m} \frac{1}{\sigma_i^2}} \displaystyle{\sum_{i=1}^m} \frac{1}{\sigma_i^2} \beta_i,
\end{split}
\end{equation}
where the second equality is due to \eqref{equ:y_tilde_sum_over_i}.

Step 3: Update $\lambda$:

\begin{equation}
    \lambda^{\rm new} = \lambda^{\rm old} + \rho \left( \frac{1}{\displaystyle{\sum_{i=1}^m} \frac{1}{\sigma_i^2}} \displaystyle{\sum_{i=1}^m} \frac{1}{\sigma_i^2} \beta_i - \delta_0 \right)
\end{equation}

The model fitting algorithm is described in Algorithm~\ref{tab_ADMM}.

\begin{algorithm*}
\caption{Alternating direction method of multipliers}
\label{tab_ADMM}
\begin{algorithmic}[1]
\Require Log-transformed gene expression measurements: $\{\{y_{ij}\}_{i=1}^m\}_{j=1}^n$, predictor variables: $\{x_j\}_{j=1}^n$ and estimated noise variance: $\{\sigma_i^2\}_{i=1}^m$.
\State Transform data. Normalize $\{x_j\}_{j=1}^n$ to zero mean and unit norm:
\begin{equation*}
    \tilde{x}_j \leftarrow \frac{x_j-\bar{x}}{\sqrt{\displaystyle{\sum_{j=1}^n} \left(x_j-\bar{x}\right)^2}}, \mbox{ with } \bar{x} \coloneqq \frac{1}{n} \displaystyle{\sum_{j=1}^n} x_j.
\end{equation*}
Center $y_{ij}$ to zero mean over row index $i$ and column index $j$: calculate $\tilde{y}_{ij}$ according to \eqref{expression:tilde_y_ij}.

\State \textit{Initialization}: Set $\rho>0$ to any fixed constant, e.g., $\rho=1$ \cite{Boyd2011}; choose the penalty parameter $\alpha$ according to Section \ref{subsec:estimate_penalty_hyperparameter}.
\State $k=0$; randomly initialize $\beta=\beta_i^0$, $i=1,2,\dots,m$, $\delta_0=\delta_0^0$, and $\lambda=\lambda^0$.
\Repeat
\State Update $\beta_i$, $i=1,2,\dots,m$
\For{$i=1,2,\dots,m$}
\begin{equation}\label{equ:derivatives_with_respect_to_beta_i}
\beta_i^{k+1} = \frac{\sigma_i^2\left(\displaystyle{\sum_{\ell=1}^m} \sigma_\ell^{-2}\right)^2}{\sigma_i^2\left(\displaystyle{\sum_{\ell=1}^m} \sigma_\ell^{-2}\right)^2+\rho}
T_{\sigma_i^2\alpha} \left[ \left(\sum_{j=1}^n \tilde{x}_j \tilde{y}_{ij} + \delta_0^k \right) - \frac{\rho}{\displaystyle{\sum_{\ell=1}^m} \frac{1}{\sigma_\ell^2}} \left(\frac{1}{\displaystyle{\sum_{\ell=1}^m} \frac{1}{\sigma_\ell^2}} \displaystyle{\sum_{\ell\neq i}} \frac{1}{\sigma_\ell^2} \beta_\ell^k - \delta_0 + \frac{\lambda^k}{\rho}\right) \right].
\end{equation}
\EndFor

\State Update $\delta_0$:
\begin{equation}
\delta_0^{k+1} = \frac{1}{\displaystyle{\sum_{i=1}^m} \frac{1}{\sigma_i^2}+\rho} \lambda^k + \frac{1}{\displaystyle{\sum_{i=1}^m} \frac{1}{\sigma_i^2}} \displaystyle{\sum_{i=1}^m} \frac{1}{\sigma_i^2} \beta_i^{k+1}.
\end{equation}

\State Update $\lambda$:
\begin{equation}\label{}
\lambda^{k+1} = \lambda^k + \rho \left( \frac{1}{\displaystyle{\sum_{i=1}^m} \frac{1}{\sigma_i^2}} \displaystyle{\sum_{i=1}^m} \frac{1}{\sigma_i^2} \beta_i^{k+1} - \delta_0^{k+1} \right)
\end{equation}
\State $k\gets k+1$;
\Until{convergence or maximum number of iterations is reached.}
\Ensure $\beta_i=\beta_i^k$, $i=1,2,\dots,m$, $\bar{\beta}^{(w)}=\delta_0^k$, and
\begin{equation}
    \beta_{0i} = \bar{y}_{i\cdot} + \bar{y}_{\cdot 1}^{(w)} - \bar{y}^{(w)} - \tilde{x}_1 \bar{\beta}^{(w)}, \; i=1,2,\dots,m
\end{equation}
\begin{equation}
    d_1 = 0, \quad d_j = \left(\bar{y}_{\cdot j}^{(w)}-\bar{y}_{\cdot 1}^{(w)}\right) - \left(\tilde{x}_j-\tilde{x}_1\right) \bar{\beta}^{(w)}, \; j=2,\dots,n
\end{equation}

\State Recover the original parameter space:
\begin{equation}
    \beta_{0i}'=\beta_{0i}-\frac{\beta_i\bar{x}}{\sqrt{\displaystyle{\sum_{j=1}^n} \left(x_j-\bar{x}\right)^2}}, \quad \beta_i'=\frac{\beta_i}{\sqrt{\displaystyle{\sum_{j=1}^n} \left(x_j-\bar{x}\right)^2}}, \quad i=1,2,\dots,m.
\end{equation}
\end{algorithmic}
\end{algorithm*}

\subsection{Estimation of Penalty Parameter $\alpha$}
\label{subsec:estimate_penalty_hyperparameter}

The \eqref{equ:objective_function_simplfied} can be expressed in matrix form as
\begin{equation}\label{equ:objective_function_simplfied_matrix_form}
f\left(\mat{\beta}\right) = \frac{1}{2} \fronorm{\mat{\Sigma}^{1/2} \left( \mat{\widetilde{Y}} - \mat{M} \mat{\beta} \mat{x}^\trans \right)}^2 + \alpha \normof{\mat{\beta}}{1},
\end{equation}
where
\begin{equation}\label{expression:Sigma}
    \mat{\Sigma} = \diag\{\mat{\sigma}\},
\end{equation}
with
\begin{equation}\label{expression:sigma}
    \mat{\sigma} = \left(
                                        \begin{array}{cccc}
                                          1/\sigma_1^2 & 1/\sigma_2^2 & \cdots & 1/\sigma_m^2 \\
                                        \end{array}
                                      \right)^\trans,
\end{equation}
and
\begin{equation}\label{expression:M}
\begin{split}
    \mat{M} &=
              \left(
                                             \begin{array}{cccc}
                                               1 & 0 & \cdots & 0 \\
                                               0 & 1 & \cdots & 0 \\
                                               0 & 0 & \ddots & 0 \\
                                               0 & 0 & \cdots & 1 \\
                                             \end{array}
                                           \right) -
                                 \frac{1}{\displaystyle{\sum_{i=1}^m} \frac{1}{\sigma_i^2}} \left(
                                   \begin{array}{cccc}
                                     1/\sigma_1^2 & 1/\sigma_2^2 & \cdots & 1/\sigma_m^2 \\
                                     1/\sigma_1^2 & 1/\sigma_2^2 & \cdots & 1/\sigma_m^2 \\
                                     \vdots & \vdots & \ddots & \vdots \\
                                     1/\sigma_1^2 & 1/\sigma_2^2 & \cdots & 1/\sigma_m^2 \\
                                   \end{array}
                                 \right)\\
    &= \mat{I}_m - \frac{1}{\displaystyle{\sum_{i=1}^m} \frac{1}{\sigma_i^2}} \mat{1}_m \mat{\sigma}^\trans.
\end{split}
\end{equation}

After expansion, \eqref{equ:objective_function_simplfied_matrix_form} becomes
\begin{equation}\label{equ:objective_function_simplfied_vector_form_expanded}
f\left(\mat{\beta}\right) = \frac{1}{2} \fronorm{\mat{\Sigma}^{1/2} \mat{\widetilde{Y}}}^2 - \mat{\beta}^\trans \mat{M}^\trans \mat{\Sigma} \mat{\widetilde{Y}} \mat{x} + \frac{1}{2} \mat{\beta}^\trans \mat{M}^\trans \mat{\Sigma} \mat{M} \mat{\beta} + \alpha \normof{\mat{\beta}}{1},
\end{equation}
where we exploit the assumption $\mat{x}^\trans \mat{x} = 1$.

Since $\frac{1}{2} \mat{\beta}^\trans \mat{M}^\trans \mat{\Sigma} \mat{M} \mat{\beta}\geq 0$ with equality occurring at $\mat{\beta}=\mat{0}$, it is shown that $\mat{\hat{\beta}} = \mat{0}$ is the minimizer of $f\left(\mat{\beta}\right)$ when
\begin{equation}\label{equ:simple_regression_model_upper_bound_of_alpha}
     \alpha \geq \normof{\mat{M}^\trans \mat{\Sigma} \mat{\widetilde{Y}} \mat{x}}{\infty} \coloneqq \max_{1\leq i \leq m} \abs{{\mat{m}^i}^\trans \mat{\Sigma} \mat{\widetilde{Y}} \mat{x}},
\end{equation}
where $\mat{m}^i$ denotes the $i$-th column of $\mat{M}$ in \eqref{expression:M}.

Note that
\begin{equation}\label{equ:M_trans_Y_tilde_simplfied}
     \mat{M}^\trans \mat{\Sigma} \mat{\widetilde{Y}} = \left(\mat{I}_m - \frac{1}{\displaystyle{\sum_{i=1}^m} \frac{1}{\sigma_i^2}} \mat{\sigma} \mat{1}_m^\trans \right) \mat{\Sigma} \mat{\widetilde{Y}} = \mat{\Sigma} \mat{\widetilde{Y}},
\end{equation}
where the last equality holds because $\mat{1}_m^\trans \mat{\Sigma} \mat{\widetilde{Y}} = \mat{0}$ due to \eqref{equ:y_tilde_sum_over_i}.

Substituting \eqref{equ:M_trans_Y_tilde_simplfied} into \eqref{equ:simple_regression_model_upper_bound_of_alpha} yields
\begin{equation}\label{equ:simple_regression_model_upper_bound_of_alpha_simplified}
     \alpha_{\rm max} = \normof{\mat{\Sigma} \mat{\widetilde{Y}} \mat{x}}{\infty} = \max_j \abs{ \frac{1}{\sigma_i^2} \mat{x}^\trans \mat{\tilde{y}}_i }.
\end{equation}

Our strategy is to set $\alpha=\epsilon \alpha_{\rm max}$, where $0<\epsilon<1$. Empirically we found that $\epsilon \in [0.001,0.1]$ works well in a wide range of parameter settings. We set $\epsilon=0.01$ in Section~\ref{sec:experiments}.

\subsection{Estimation of $\{\sigma_i^2\}_{i=1}^m$}
\label{subsec:estimate_noise_variance}

To solve for $\{\sigma_i^2\}_{i=1}^m$, consider the negative log-likelihood function in \eqref{equ:likelihood} with $\{\sigma_i^2\}_{i=1}^m$ being unknown parameters as well:
\begin{equation}\label{equ:negative_log-likelihood_full}
l = \displaystyle{\sum_{i=1}^m} \left[ \frac{n}{2} \log(2\pi\sigma_i^2) + \frac{1}{2\sigma_i^2}
 \displaystyle{\sum_{j=1}^n} \left(y_{ij}-\beta_{0i}-x_j \beta_i-d_j\right)^2 \right].
\end{equation}

Taking partial derivatives of $l(\cdot)$ with respect to $d_j$ and $\beta_{0i}$ and setting the results to zero, we arrive at \eqref{equ:d_j} and \eqref{equ:mu_i_no_d_j_simplfied} respectively. The sum of \eqref{equ:d_j} and \eqref{equ:mu_i_no_d_j_simplfied} gives \eqref{equ:mu_i_plus_d_j}.

Taking partial derivatives of $l(\cdot)$ with respect to $\beta_i$ and setting the result to zero, we have
\begin{equation}\label{equ:beta_i}
    \beta_i = \displaystyle{\sum_{j=1}^n} x_j y_{ij} - \displaystyle{\sum_{j=1}^n} x_j \left( \beta_{0i}+d_j \right).
\end{equation}
Substituting \eqref{equ:mu_i_plus_d_j} into \eqref{equ:beta_i} yields
\begin{equation}\label{equ:beta_i_original}
    \beta_i = \displaystyle{\sum_{j=1}^n} x_j y_{ij} - \frac{1}{\displaystyle{\sum_{i=1}^m} \frac{1}{\sigma_i^2}} \displaystyle{\sum_{i=1}^m} \frac{1}{\sigma_i^2} \displaystyle{\sum_{j=1}^n} x_j y_{ij} + \bar{\beta}^{(w)},
\end{equation}
where $\bar{\beta}^{(w)}$ is defined in \eqref{equ:weighted_average_beta_over_i}.

Taking partial derivatives of $\sigma_i^2$ and setting the result to zero gives
\begin{equation}\label{equ:sigma_i_squared}
    \sigma_i^2 = \frac{1}{n} \displaystyle{\sum_{j=1}^n} \left(y_{ij}-\beta_{0i}-x_j \beta_i-d_j\right)^2.
\end{equation}
Substituting \eqref{equ:mu_i_plus_d_j} into \eqref{equ:sigma_i_squared} yields
\begin{equation}\label{equ:sigma_i_squared_original}
    \sigma_i^2 = \frac{1}{n} \displaystyle{\sum_{j=1}^n} \left(y_{ij} - \bar{y}_{i\cdot} - \bar{y}_{\cdot j}^{(w)} + \bar{y}^{(w)} - x_j \beta_i + x_j \bar{\beta}^{(w)} \right)^2,
\end{equation}
where $\bar{y}_{i\cdot}$, $\bar{y}_{\cdot j}^{(w)}$ and $\bar{y}^{(w)}$ are defined in \eqref{equ:average_y_over_j}, \eqref{equ:weighted_average_y_over_i} and \eqref{equ:weighted_average_y_over_ij}, respectively.

Given initial estimates for $\bar{\beta}^{(w)}$ and $\{\sigma_i^2\}_{i=1}^m$, we can alternate equations \eqref{equ:beta_i_original}, \eqref{equ:sigma_i_squared_original} and \eqref{equ:weighted_average_beta_over_i} iteratively to graduately refine the estimates for $\beta_i$ and $\sigma_i^2$, as shown in Algorithm \ref{tab_estimation_noise_variance}.

\begin{algorithm}
\caption{Estimation of $\{\sigma_i^2\}_{i=1}^m$}
\label{tab_estimation_noise_variance}
\begin{algorithmic}[1]
\Require Log-transformed gene expression measurements: $\{\{y_{ij}\}_{i=1}^m\}_{j=1}^n$ and predictor variables: $\{x_j\}_{j=1}^n$.
\State Normalize $\{x_j\}_{j=1}^n$ to zero mean and unit norm:
\begin{equation*}
    x_j \leftarrow \frac{x_j-\bar{x}}{\sqrt{\displaystyle{\sum_{j=1}^n} \left(x_j-\bar{x}\right)^2}}, \mbox{ with } \bar{x} \coloneqq \frac{1}{n} \displaystyle{\sum_{j=1}^n} x_j.
\end{equation*}
\State \textit{Initialization}: $\bar{\beta}^{(w)} = 0$, $\sigma_1^2 = \sigma_2^2 = \dotsb = \sigma_m^2 = 1$.
\Repeat
\For{$i=1,2,\dots,m$}

Update $\beta_i$, $i=1,2,\dots,m$ according to \eqref{equ:beta_i_original}.
\EndFor

\State Update $\{\sigma_i^2\}_{i=1}^m$ according to \eqref{equ:sigma_i_squared_original}.

\State Update $\bar{\beta}^{(w)}$ according to \eqref{equ:weighted_average_beta_over_i}.

\Until{convergence or maximum number of iterations is reached.}
\Ensure $\hat{\sigma}_i^2 = \sigma_i^2$, $i=1,2,\dots,m$.
\end{algorithmic}
\end{algorithm}

Then we take another weighted average of $\hat{\sigma}_i^2$ and the estimated mean variance across all the genes to obtain a robust estimate for $\sigma_i^2$. That is
\begin{equation}\label{equ:Bayes_shrinkage_estimator_of_noise_variance}
    \hat{\sigma}_i'^2=(1-w)\hat{\sigma}_i^2+w\overline{\hat{\sigma}^2}
\end{equation}
where
\begin{equation}
    \overline{\hat{\sigma}^2}=\frac{1}{m} \sum_{i=1}^m \hat{\sigma}_i^2,
\end{equation}
and the weight $w$ is calculated using the following formula as suggested in~\cite{Ji2005} which is based on an empirical Bayes approach
\begin{equation}
    w=\frac{2(m-1)}{n+1}\left(\frac1m+\frac{(\overline{\hat{\sigma}^2})^2}{\sum_{i=1}^m\left(\hat{\sigma}_i^2-\overline{\hat{\sigma}^2}\right)^2}\right).
\end{equation}
This kind of variance estimation approach is widely used in differential gene expression analysis with small sample sizes \cite{Ji2010,Smyth2004}. The estimated variances $\hat{\sigma}_i'^2$, $i=1,2,\dots,m$, can then be used in Algorithm \ref{tab_ADMM} to solve for $\{\beta_i\}_{i=1}^m$.

\rev{\begin{remark}
In the special case of $ \sigma_1^2 = \sigma_2^2 = \dotsb = \sigma_m^2 = \sigma^2 $, it no longer requires to estimate $ \sigma^2$ since the unknown $\sigma^2$ in \eqref{equ:objective_function} can be absorbed into the penalty parameter $\alpha$.
\end{remark}}

\section{Extension to Multiple Linear Regression Model and Algorithm Development}
\label{sec:multiple_regression_model}

\begin{figure*}[!b]
\normalsize
\vspace*{4pt}
\hrulefill
\setcounter{MYtempeqncnt}{\value{equation}}
\setcounter{equation}{64}
\begin{equation}\label{equ:mlr_model_1_objective_function_augmented_Lagrangian}
L_\rho\left(\{\mat{\beta}_i\}, \mat{\delta}_0, \mat{\lambda}\right) = \displaystyle{\sum_{i=1}^m} \frac{1}{2\sigma_i^2} \displaystyle{\sum_{j=1}^n} \left(\tilde{y}_{ij} - \mat{x}_j^\trans \mat{\beta}_i + \mat{x}_j^\trans \mat{\delta}_0\right)^2 + \alpha \displaystyle{\sum_{i=1}^m} \, \abs{\beta_{ip}}
+ \mat{\lambda}^\trans \left(\frac{1}{\displaystyle{\sum_{i=1}^m} \frac{1}{\sigma_i^2}} \displaystyle{\sum_{i=1}^m} \frac{1}{\sigma_i^2} \mat{\beta}_i - \mat{\delta}_0\right)
+ \frac{\rho}{2} \twonorm{\frac{1}{\displaystyle{\sum_{i=1}^m} \frac{1}{\sigma_i^2}} \displaystyle{\sum_{i=1}^m} \frac{1}{\sigma_i^2} \mat{\beta}_i - \mat{\delta}_0}^2.
\end{equation}
\setcounter{equation}{\value{MYtempeqncnt}}
\end{figure*}


In the multiple linear regression model, each response or outcome is modeled by $p>1$ predictors:
\begin{equation}\label{equ:multiple_regression_model}
    y_{ij} \sim \mathcal{N}\left(\beta_{0i}+\mat{\beta}_i^\trans \mat{x}_j + d_j,\sigma_i^2\right)
\end{equation}
where
\begin{equation}\label{equ:mlr_model_beta_i}
    \mat{\beta}_i = \left(
                      \begin{array}{c}
                        \beta_{i1} \\
                       \beta_{i2} \\
                        \vdots \\
                        \beta_{ip} \\
                      \end{array}
                    \right) \in \real^{p \times 1}
\end{equation}
is a vector of regression coefficients representing log-fold-change of expression levels of gene $i$ between treatment conditions, and
\begin{equation}\label{equ:mlr_model_x_j}
    \mat{x}_j = \left(
                  \begin{array}{c}
                    x_{j1} \\
                    x_{j2} \\
                    \vdots \\
                    x_{jp} \\
                  \end{array}
                \right) \in \real^{p \times 1}
\end{equation}
is a vector of independent/explanatory variables representing the treatment conditions (drug dosage, blood pressure, age, BMI, etc.) for sample $j$, and $\beta_{0i}$, $d_j$, and $\sigma_i$ respectively represent the $y$-intercept of gene $i$, scaling factor for sample $j$ and standard deviation of log-transformed expression levels of gene $i$, as defined in the simple regression model.

Since the sample values are independent across the genes and samples, the likelihood is given by
\begin{equation}\label{equ:mlr_model_likelihood}
\Prob{\mat{y} | \mat{\beta}_0, \{\mat{\beta}_i\}, \mat{d}} = \displaystyle{\prod_{i=1}^m} \displaystyle{\prod_{j=1}^n} \frac{1}{\sqrt{2\pi \sigma_i^2}} \exp\left\{ -\frac{\left(y_{ij}-\beta_{0i}-\mat{\beta}_i^\trans \mat{x}_j-d_j\right)^2}{2\sigma_i^2} \right\}.
\end{equation}


Assume that $ \{\sigma_i^2\}_{i=1}^m$ are known, maximization of \eqref{equ:mlr_model_likelihood} leads to minimizing the negative log-likelihood:
\begin{equation}\label{equ:mlr_model_negative_log-likelihood_equal_variance}
l\left(\mat{\beta}_0, \{\mat{\beta}_i\}, \mat{d}; \mat{y}\right) = \displaystyle{\sum_{i=1}^m\sum_{j=1}^n} \frac{1}{2\sigma_i^2} \left(y_{ij}-\beta_{0i}-\mat{\beta}_i^\trans \mat{x}_j-d_j\right)^2
\end{equation}

The objective function to be minimized is
\begin{equation}\label{equ:mlr_model_objective_function}
f\left(\mat{\beta}_0, \{\mat{\beta}_i\}, \mat{d}\right) = \displaystyle{\sum_{i=1}^m\sum_{j=1}^n} \frac{1}{2\sigma_i^2} \left(y_{ij}-\beta_{0i}-\mat{x}_j^\trans \mat{\beta}_i-d_j\right)^2 + \displaystyle{\sum_{i=1}^m} \,  p\left(\mat{\beta}_i\right).
\end{equation}

Below we introduce two types of penalty function $p\left(\mat{\beta}_i\right)$.

\begin{enumerate}
  \item Type I penalty:
\begin{equation}\label{equ:mlr_model_penalty_function_1}
    p\left(\mat{\beta}_i\right) = \alpha \abs{\beta_{ip}}.
\end{equation}
Gene $i$ is differentially expressed if $\beta_{ip} \neq 0$ and not otherwise. This penalty is for the applications where one covariate is of main interest (e.g., treatment) while we want to adjust for all possible effects of other confounding covariates (e.g., age, gender, etc).

  \item Type II penalty:
\begin{equation}\label{equ:mlr_model_penalty_function_2}
    p\left(\mat{\beta}_i\right) = \alpha \twonorm{\mat{\beta}_i}.%
\end{equation}
Gene $i$ is differentially expressed if $\mat{\beta}_i \neq 0$ and not otherwise. This penalty is for the applications where all covariates are of interest and we want to identify the genes for which at least one covariate has an effect.
\end{enumerate}

It can be proved that the optimization problem \eqref{equ:mlr_model_objective_function} with penalty  \eqref{equ:mlr_model_penalty_function_1} or \eqref{equ:mlr_model_penalty_function_2} is jointly convex in $\left(\mat{\beta}_0, \{\mat{\beta}_i\}, \mat{d}\right)$.

Assume that
\begin{equation}\label{equ:mlr_model_assumption}
    \displaystyle{\sum_{j=1}^n} \mat{x}_j = 0,
\end{equation}
and set $d_1=0$. Using similar argumentation as in Section \ref{subsec:simple_regression_model} to eliminate $\mat{\beta}_0$ and $\mat{d}$, we simplify \eqref{equ:mlr_model_objective_function} to
\begin{equation}\label{equ:mlr_model_objective_function_simplfied}
f\left(\{\mat{\beta}_i\}\right) = \displaystyle{\sum_{i=1}^m} \frac{1}{2\sigma_i^2}
 \displaystyle{\sum_{j=1}^n} \left(\tilde{y}_{ij} - \mat{x}_j^\trans \mat{\beta}_i + \mat{x}_j^\trans \mat{\bar{\beta}}^{(w)}\right)^2 + \displaystyle{\sum_{i=1}^m} \,  p\left(\mat{\beta}_i\right),
\end{equation}
where $\tilde{y}_{ij}$ \rev{is the same as that in} \eqref{expression:tilde_y_ij}, and
\begin{equation}\label{equ:mlr_model_weighted_average_beta_over_i}
    \mat{\bar{\beta}}^{(w)} \coloneqq \frac{1}{\displaystyle{\sum_{i=1}^m} \frac{1}{\sigma_i^2}} \displaystyle{\sum_{i=1}^m} \frac{1}{\sigma_i^2} \mat{\beta}_i.
\end{equation}


\subsection{Regression with Type I Penalty: Model Fitting by ADMM}
\label{subsec:TypeIPenalty_based_Optimization_byADMM}

\begin{figure*}[!b]
\normalsize
\vspace*{4pt}
\hrulefill
\setcounter{MYtempeqncnt}{\value{equation}}
\setcounter{equation}{75}
\begin{equation}\label{equ:mlr_model_2_objective_function_augmented_Lagrangian}
L_\rho\left(\{\mat{\beta}_i\}, \mat{\delta}_0, \mat{\lambda}\right) = \displaystyle{\sum_{i=1}^m} \frac{1}{2\sigma_i^2}
 \displaystyle{\sum_{j=1}^n} \left(\tilde{y}_{ij} - \mat{x}_j^\trans \mat{\beta}_i + \mat{x}_j^\trans \mat{\delta}_0\right)^2 + \alpha \displaystyle{\sum_{i=1}^m} \, \twonorm{\mat{\beta}_i}
 + \mat{\lambda}^\trans \left(\frac{1}{\displaystyle{\sum_{i=1}^m} \frac{1}{\sigma_i^2}} \displaystyle{\sum_{i=1}^m} \frac{1}{\sigma_i^2} \mat{\beta}_i - \mat{\delta}_0\right)
+ \frac{\rho}{2} \twonorm{\frac{1}{\displaystyle{\sum_{i=1}^m} \frac{1}{\sigma_i^2}} \displaystyle{\sum_{i=1}^m} \frac{1}{\sigma_i^2} \mat{\beta}_i - \mat{\delta}_0}^2.
\end{equation}
\setcounter{equation}{\value{MYtempeqncnt}}
\end{figure*}

\begin{figure*}[!b]
\normalsize
\vspace*{4pt}
\hrulefill
\setcounter{MYtempeqncnt}{\value{equation}}
\setcounter{equation}{76}
\begin{equation}\label{equ:mlr_model_2_objective_function_augmented_Lagrangian_i}
\begin{split}
L_i\left(\{\mat{\beta}_i\}, \mat{\delta}_0, \mat{\lambda}\right) = &\frac{1}{2\sigma_i^2} \displaystyle{\sum_{j=1}^n} \left(\tilde{y}_{ij} - \mat{x}_j^\trans \mat{\beta}_i + \mat{x}_j^\trans \mat{\delta}_0\right)^2 + \alpha \twonorm{\mat{\beta}_i}
+ \mat{\lambda}^\trans \frac{1}{\displaystyle{\sum_{\ell=1}^m} \frac{1}{\sigma_\ell^2}} \frac{1}{\sigma_i^2} \mat{\beta}_i + \frac{\rho}{2} \left(\frac{1}{\displaystyle{\sum_{\ell=1}^m} \frac{1}{\sigma_\ell^2}} \frac{1}{\sigma_i^2} \mat{\beta}_i + \frac{1}{\displaystyle{\sum_{\ell=1}^m} \frac{1}{\sigma_\ell^2}} \displaystyle{\sum_{\ell\neq i}} \frac{1}{\sigma_\ell^2} \mat{\beta}_\ell - \mat{\delta}_0\right)^2 \\
= & \frac{1}{2} \mat{\beta}_i^\trans \left( \frac{1}{\sigma_i^2} \mat{X}^\trans \mat{X} + \frac{\rho}{\sigma_i^4} \frac{1}{\left(\displaystyle{\sum_{\ell=1}^m} \frac{1}{\sigma_\ell^2}\right)^2} \mat{I}_p \right) \mat{\beta}_i - \mat{\beta}_i^\trans \mat{v}_i + \alpha \twonorm{\mat{\beta}_i} + c,
\end{split}
\end{equation}
\setcounter{equation}{\value{MYtempeqncnt}}
\end{figure*}


To apply the ADMM, we reformulate the Type I penalized regression problem as
\begin{subequations}\label{equ:mlr_model_1_objective_function_simplfied_ADMM}
\begin{equation}
f\left(\{\mat{\beta}_i\},\mat{\delta}_0\right) = \displaystyle{\sum_{i=1}^m} \frac{1}{2\sigma_i^2} \displaystyle{\sum_{j=1}^n} \left(\tilde{y}_{ij} - \mat{x}_j^\trans \mat{\beta}_i + \mat{x}_j^\trans \mat{\delta}_0\right)^2 + \alpha \displaystyle{\sum_{i=1}^m} \, \abs{\beta_{ip}},
\end{equation}
subject to
\begin{eqnarray}
  \frac{1}{\displaystyle{\sum_{i=1}^m} \frac{1}{\sigma_i^2}} \displaystyle{\sum_{i=1}^m} \frac{1}{\sigma_i^2} \mat{\beta}_i &=& \mat{\delta}_0.
\end{eqnarray}
\end{subequations}

The augmented Lagrangian of \eqref{equ:mlr_model_1_objective_function_simplfied_ADMM} is \eqref{equ:mlr_model_1_objective_function_augmented_Lagrangian} \rev{at the bottom of the page.} \addtocounter{equation}{1}

Step 1: Update $\mat{\beta}_i$, $i=1,2,\dots,m$:


Taking partial derivative of \eqref{equ:mlr_model_1_objective_function_augmented_Lagrangian} with respect to $\mat{\beta}_i$ and setting the result to zero gives
%
\begin{equation}\label{equ:mlr_model_1_derivatives_with_respect_to_beta_i}
\left[ \frac{1}{\sigma_i^2} \mat{X}^\trans \mat{X} + \frac{\rho}{\sigma_i^4} \frac{1}{\left(\displaystyle{\sum_{\ell=1}^m} \frac{1}{\sigma_\ell^2}\right)^2} \mat{I}_p \right] \mat{\beta}_i + \alpha \left(
                  \begin{array}{c}
                    0 \\
                    \vdots \\
                    0 \\
                    \partial \abs{\beta_{ip}} \\
                  \end{array}
                \right) = \mat{v}_i.
\end{equation}
where
\begin{equation}\label{equ:def_of_X}
    \mat{X} = \left(
                \begin{array}{c}
                  \mat{x}_1^\trans \\
                  \mat{x}_2^\trans \\
                  \vdots \\
                  \mat{x}_n^\trans \\
                \end{array}
              \right)
            = \left(
                \begin{array}{cccc}
                  x_{11} & x_{12} & \cdots & x_{1p} \\
                  x_{21} & x_{22} & \cdots & x_{2p} \\
                  \vdots & \vdots & \ddots & \vdots \\
                  x_{n1} & x_{n2} & \cdots & x_{np} \\
                \end{array}
              \right)
            \in \real^{n \times p},
\end{equation}
$\partial \abs{\beta_{ip}}$ is the subgradient of $\abs{\beta_{ip}}$ with respect to $\beta_{ip}$, and
\begin{equation}\label{equ:def_of_vi}
\begin{split}
    \mat{v}_i = &\frac{1}{\sigma_i^2} \left(\sum_{j=1}^n \mat{x}_j \tilde{y}_{ij} + \mat{X}^\trans \mat{X} \mat{\delta}_0 \right) \\
    &- \frac{\rho}{\sigma_i^2} \frac{1}{\displaystyle{\sum_{\ell=1}^m} \frac{1}{\sigma_\ell^2}}  \left(\frac{1}{\displaystyle{\sum_{\ell=1}^m} \frac{1}{\sigma_\ell^2}} \displaystyle{\sum_{\ell\neq i}} \frac{1}{\sigma_\ell^2} \mat{\beta}_\ell - \mat{\delta}_0 + \frac{\mat{\lambda}}{\rho}\right),
\end{split}
\end{equation}


\rev{Using the block matrix representation:
\begin{equation*}
    \frac{1}{\sigma_i^2} \mat{X}^\trans \mat{X} + \frac{\rho}{\sigma_i^4} \frac{1}{\left(\displaystyle{\sum_{\ell=1}^m} \frac{1}{\sigma_\ell^2}\right)^2} \mat{I}_p = \mat{Q} = \left(
                                                                      \begin{array}{cc}
                                                                        \mat{\widetilde{Q}} & \mat{q} \\
                                                                        \mat{q}^\trans & q_{pp} \\
                                                                      \end{array}
                                                                    \right), \quad
\end{equation*}
\begin{equation*}
    \mat{\beta}_i = \left(
                    \begin{array}{c}
                      \mat{\beta}_i^{-} \\
                      \beta_{ip} \\
                    \end{array}
                  \right), \quad
    \mat{v}_i = \left(
                \begin{array}{c}
                  \mat{v}_i^{-} \\
                  v_{ip} \\
                \end{array}
              \right),
\end{equation*}
where $\mat{\widetilde{Q}}$ is the submatrix of $\mat{Q}$ with last row and last column deleted, from \eqref{equ:mlr_model_1_derivatives_with_respect_to_beta_i}} we have
\begin{eqnarray}
  \mat{\widetilde{Q}}\mat{\beta}_i^{-} + \mat{q} \beta_{ip} &=& \mat{v}_i^{-} \label{equ:Q_minus} \\
  \mat{q}^\trans \mat{\beta}_i^{-} + q_{pp} \beta_{ip} + \alpha \partial \abs{\beta_{ip}} &=& v_{ip} \label{equ:q_p}.
\end{eqnarray}

From \eqref{equ:Q_minus} it follows
\begin{equation}\label{equ:beta_minus}
    \mat{\beta}_i^{-} = \mat{\widetilde{Q}}^{-1} \left( \mat{v}_i^{-} - \mat{q} \beta_{ip} \right).
\end{equation}

Substituting \eqref{equ:beta_minus} into \eqref{equ:q_p} yields
\begin{equation}\label{equ:beta_p}
    \beta_{ip} = \frac{1}{q_{pp}-\mat{q}^\trans \mat{\widetilde{Q}}^{-1} \mat{q} } T_\alpha \left[ v_{ip} - \mat{q}^\trans \mat{\widetilde{Q}}^{-1} \mat{v}_i^{-}\right].
\end{equation}

Step 2: Update $\mat{\delta}_0$:

%

Taking the derivative of \eqref{equ:mlr_model_1_objective_function_augmented_Lagrangian} with respect to $\mat{\delta}_0$ and setting the result to zero gives
\begin{equation}\label{equ:mlr_delta_0}
\mat{\delta}_0 = \left(\displaystyle{\sum_{i=1}^m} \frac{1}{\sigma_i^2} \mat{X}^\trans \mat{X}+\rho\mat{I}_p\right)^{-1} \mat{\lambda} + \frac{1}{\displaystyle{\sum_{i=1}^m} \frac{1}{\sigma_i^2}} \displaystyle{\sum_{i=1}^m} \frac{1}{\sigma_i^2} \mat{\beta}_i,
\end{equation}
where we have exploited \eqref{equ:y_tilde_sum_over_i}.

Step 3: Update $\mat{\lambda}$:
\begin{equation}
\mat{\lambda}^{\rm new} = \mat{\lambda}^{\rm old} + \rho \left( \frac{1}{\displaystyle{\sum_{i=1}^m} \frac{1}{\sigma_i^2}} \displaystyle{\sum_{i=1}^m} \frac{1}{\sigma_i^2} \mat{\beta}_i - \mat{\delta}_0\right).
\end{equation}

\subsection{Regression with Type II Penalty: Model Fitting by ADMM}
\label{subsec:TypeIIPenalty_based_Optimization_byADMM}

\rev{The Type II penalized regression problem is reformulated as
\begin{subequations}\label{equ:mlr_model_2_objective_function_simplfied_ADMM}
\begin{equation}
f\left(\{\mat{\beta}_i\},\mat{\delta}_0\right) = \displaystyle{\sum_{i=1}^m} \frac{1}{2\sigma_i^2}
 \displaystyle{\sum_{j=1}^n} \left(\tilde{y}_{ij} - \mat{x}_j^\trans \mat{\beta}_i + \mat{x}_j^\trans \mat{\delta}_0\right)^2 + \alpha \displaystyle{\sum_{i=1}^m} \, \twonorm{\mat{\beta}_i},
\end{equation}
subject to
\begin{eqnarray}
  \frac{1}{\displaystyle{\sum_{i=1}^m} \frac{1}{\sigma_i^2}} \displaystyle{\sum_{i=1}^m} \frac{1}{\sigma_i^2} \mat{\beta}_i &=& \mat{\delta}_0.
\end{eqnarray}
\end{subequations}}

The augmented Lagrangian of \eqref{equ:mlr_model_2_objective_function_simplfied_ADMM} is \eqref{equ:mlr_model_2_objective_function_augmented_Lagrangian} \rev{at the bottom of the page.} \addtocounter{equation}{1}

Step 1: Update $\mat{\beta}_i$, $i=1,2,\dots,m$:

The relevant terms to compute the derivatives of \eqref{equ:mlr_model_2_objective_function_augmented_Lagrangian} with respect to $\mat{\beta}_i$ is \eqref{equ:mlr_model_2_objective_function_augmented_Lagrangian_i} \rev{at the bottom of the page}, \addtocounter{equation}{1}
where $c$ is an irrelevant constant which does not depend on $\mat{\beta}_i$, and $\mat{v}_i$ is defined in \eqref{equ:def_of_vi}.

It can be shown when $\twonorm{\mat{v}_i} \leq \alpha$ then $\mat{\beta}_i = \mat{0}$; otherwise denote the eigendecomposition of $\mat{X}^\trans \mat{X}$ as $\mat{X}^\trans \mat{X} = \mat{U} \mat{D} \mat{U}^\trans$, we have that minimization of \eqref{equ:mlr_model_2_objective_function_augmented_Lagrangian_i} is equivalent to
\begin{subequations}\label{equ:mlr_model_2_equation_update_beta_i}
\begin{equation}
    \min_{\mat{\beta}_i} \frac{1}{2} \twonorm{ \mat{Z}\mat{\beta}_i- \mat{b}_i }^2 + \alpha \twonorm{\mat{\beta}_i}.
\end{equation}
where
\begin{eqnarray}
  \mat{Z} &=& \left[\frac{1}{\sigma_i^2} \mat{D} + \frac{\rho}{\sigma_i^4} \frac{1}{\left(\displaystyle{\sum_{\ell=1}^m} \frac{1}{\sigma_\ell^2}\right)^2} \mat{I}_p\right]^{1/2} \mat{U}^\trans, \\
  \mat{b}_i &=& \left[\frac{1}{\sigma_i^2} \mat{D} + \frac{\rho}{\sigma_i^4} \frac{1}{\left(\displaystyle{\sum_{\ell=1}^m} \frac{1}{\sigma_\ell^2}\right)^2} \mat{I}_p\right]^{-1/2} \mat{U}^\trans \mat{v}_i.
\end{eqnarray}
\end{subequations}

As in~\cite{Friedman2010}, we use a coordinate descent procedure to optimize \eqref{equ:mlr_model_2_equation_update_beta_i}. For each $s$, given the estimate of $\{\hat{\beta}_{i \ell}\}_{\ell\neq s}$, $\beta_{is}$ can be estimated by solving
\begin{equation}\label{equ:mlr_model_2_equation_for_solving_beta_is}
    \min_{\beta_{is}} \frac{1}{2} \twonorm{ \mat{z}_s \beta_{is} - \mat{r}_i^{(s)} }^2 + \alpha \sqrt{ \displaystyle{\beta_{is}^2 + \sum_{\ell\neq s}} \hat{\beta}_{i \ell}^2 },
\end{equation}
where
\begin{equation}
    \mat{r}_i^{(s)} = \mat{b}_i - \displaystyle{\sum_{\ell\neq s}} \mat{z}_\ell \beta_{i \ell}.
\end{equation}

We solve \eqref{equ:mlr_model_2_equation_for_solving_beta_is} via a one-dimensional search. Note that the solution to \eqref{equ:mlr_model_2_equation_for_solving_beta_is} falls between 0 and $\beta_{i \ell}^{\rm o}= \mat{z}_s^\trans \mat{r}_i^{(s)}/\twonorm{\mat{z}_s}^2$, the ordinary least-squares estimate. We can the {\tt optimize} function in the R package, or {\tt fminbnd} function in MATLAB, which perform one-dimensional search based on golden section search and successive parabolic interpolation.



After updating $\{\mat{\beta}_i\}_{i=1}^m$, the updates of $\mat{\delta}_0$ and $\mat{\lambda}$ turn out to be the same as that in Section \ref{subsec:TypeIPenalty_based_Optimization_byADMM}.

\subsection{Estimation of Penalty Parameter $\alpha$}

The \eqref{equ:mlr_model_objective_function_simplfied} can be expressed in matrix form as
\begin{equation}\label{equ:mlr_model_objective_function_simplfied_matrix_form}
f\left(\mat{B}\right) = \frac{1}{2} \fronorm{\mat{\Sigma}^{1/2} \left( \mat{\widetilde{Y}} - \mat{M} \mat{B} \mat{X}^\trans\right)}^2 + p\left(\mat{B}\right),
\end{equation}
where $\mat{M}$ and $\mat{X}$ are respectively defined in \eqref{expression:M} and \eqref{equ:def_of_X}, and
\begin{equation}\label{equ:def_of_B}
    \mat{B} = \left(
                \begin{array}{c}
                  \mat{\beta}_1^\trans \\
                  \mat{\beta}_2^\trans \\
                  \vdots \\
                  \mat{\beta}_m^\trans \\
                \end{array}
              \right)
            = \left(
                    \begin{array}{cccc}
                      \beta_{11} & \beta_{12} & \cdots & \beta_{1p} \\
                      \beta_{21} & \beta_{22} & \cdots & \beta_{2p} \\
                      \vdots & \vdots & \ddots & \vdots \\
                      \beta_{m1} & \beta_{m2} & \cdots & \beta_{mp} \\
                    \end{array}
                  \right) \in \real^{m \times p},
\end{equation}
and $p\left(\mat{B}\right)$ is the penalty function.

The derivative of $f\left(\mat{B}\right)$ with respect to $\mat{B}$ is
\begin{equation}\label{equ:mlr_model_objective_function_simplfied_matrix_form_derivative}
\frac{\partial f}{\partial \mat{B}} = \mat{M}^\trans \mat{\Sigma} \mat{M} \mat{B} \mat{X}^\trans\mat{X} - \mat{M}^\trans \mat{\Sigma} \mat{\widetilde{Y}} \mat{X} + \frac{\partial p\left(\mat{B}\right)}{\partial \mat{B}}.
\end{equation}

\subsubsection{Type I Penalty}
When $p\left(\mat{B} \right) = \alpha \displaystyle{\sum_{i=1}^m} \, \abs{\beta_{ip}}$, its derivative with respect to $\mat{B}$ is
\begin{equation}
\frac{\partial p\left(\mat{B}\right)}{\partial \mat{B}} = \alpha \left(
                                                             \begin{array}{cccc}
                                                               0 & \cdots & 0 & \partial \abs{\beta_{1p}} \\
                                                               0 & \cdots & 0 & \partial \abs{\beta_{2p}} \\
                                                               \vdots & \ddots & \vdots & \vdots \\
                                                               0 & \cdots & 0 & \partial \abs{\beta_{mp}} \\
                                                             \end{array}
                                                           \right)
                                                         = \left(
                                                             \begin{array}{cc}
                                                               \mat{0}_{m \times (p-1)} & \alpha \frac{\partial \normof{\mat{\beta}^p}{1}}{\partial {\mat{\beta}^p}} \\
                                                             \end{array}
                                                           \right).
\end{equation}

Denote
\begin{equation*}
    \mat{X} = \left(
                       \begin{array}{cccc}
                         \mat{x}^1 & \cdots & \mat{x}^{p-1} & \mat{x}^p \\
                       \end{array}
                     \right)
                   = \left(
                       \begin{array}{cc}
                         \mat{X}_1 & \mat{x}^p \\
                       \end{array}
                     \right), \quad
\end{equation*}
\begin{equation*}
    \mat{B} = \left(
                       \begin{array}{cccc}
                         \mat{\beta}^1 & \cdots & \mat{\beta}^{p-1} & \mat{\beta}^p \\
                       \end{array}
                     \right)
                   = \left(
                       \begin{array}{cc}
                         \mat{B}_1 & \mat{\beta}^p \\
                       \end{array}
                     \right).
\end{equation*}
Setting \eqref{equ:mlr_model_objective_function_simplfied_matrix_form_derivative} equal to zero gives
\begin{equation}
    \mat{M}^\trans \mat{\Sigma} \mat{M} \left( \mat{B}_1 \mat{X}_1^\trans + \mat{\beta}^p {\mat{x}^p}^\trans \right) \mat{X}_1 = \mat{M}^\trans \mat{\Sigma} \mat{\widetilde{Y}} \mat{X}_1 \label{equ:mlr_model_objective_function_simplfied_matrix_form_derivative_equation_1}
\end{equation}
\begin{equation}
    \mat{M}^\trans \mat{\Sigma} \mat{M} \left( \mat{B}_1 \mat{X}_1^\trans + \mat{\beta}^p {\mat{x}^p}^\trans \right) \mat{x}^p + \alpha \frac{\partial \normof{\mat{\beta}^p}{1}}{\partial {\mat{\beta}^p}} = \mat{M}^\trans \mat{\Sigma} \mat{\widetilde{Y}} \mat{x}^p.\label{equ:mlr_model_objective_function_simplfied_matrix_form_derivative_equation_2}
\end{equation}

Since $\mat{M}^\trans \mat{\Sigma} \mat{M}$ is rank deficient\footnote[2]{Simple calculation shows that the rank of $\mat{M}^\trans \mat{\Sigma} \mat{M}$ is $m-1$.}
, the solution to \eqref{equ:mlr_model_objective_function_simplfied_matrix_form_derivative_equation_1} is not unique. We apply the pseudoinverse of $\mat{M}^\trans \mat{\Sigma} \mat{M}$ to obtain the minimum-norm solution to \eqref{equ:mlr_model_objective_function_simplfied_matrix_form_derivative_equation_1}:
\begin{equation}\label{equ:mlr_model_objective_function_simplfied_matrix_form_derivative_equation_1_solution}
    \mat{B}_1 = \left(\mat{M}^\trans \mat{\Sigma} \mat{M}\right)^{\pinv} \left( \mat{M}^\trans \mat{\Sigma} \mat{\widetilde{Y}} - \mat{M}^\trans \mat{\Sigma} \mat{M} \mat{\beta}^p {\mat{x}^p}^\trans  \right) \mat{X}_1 \left(\mat{X}_1^\trans \mat{X}_1\right)^{-1}.
\end{equation}

Substituting \eqref{equ:mlr_model_objective_function_simplfied_matrix_form_derivative_equation_1_solution} into \eqref{equ:mlr_model_objective_function_simplfied_matrix_form_derivative_equation_2} yields
\begin{equation}\label{equ:mlr_model_objective_function_simplfied_matrix_form_derivative_equation_2_solution}
\begin{split}
     &\mat{M}^\trans \mat{\Sigma} \mat{M} \mat{\beta}^p {\mat{x}^p}^\trans \left[ \mat{I}_n - \mat{X}_1 \left(\mat{X}_1^\trans \mat{X}_1\right)^{-1} \mat{X}_1^\trans \right] {\mat{x}^p} + \alpha \frac{\partial \normof{\mat{\beta}^p}{1}}{\partial {\mat{\beta}^p}} \\
     = &\mat{M}^\trans \mat{\Sigma} \mat{\widetilde{Y}} \left[ \mat{I}_n - \mat{X}_1 \left(\mat{X}_1^\trans \mat{X}_1\right)^{-1} \mat{X}_1^\trans \right] \mat{x}^p.
\end{split}
\end{equation}
\rev{Note that to arrive at \eqref{equ:mlr_model_objective_function_simplfied_matrix_form_derivative_equation_2_solution}, we have exploited the fact that $\left(\mat{M}^\trans \mat{\Sigma} \mat{M}\right) \left(\mat{M}^\trans \mat{\Sigma} \mat{M}\right)^{\pinv} \mat{M}^\trans \mat{\Sigma} = \mat{M}^\trans \mat{\Sigma}$ which is due to that $ \mat{M}^\trans \mat{\Sigma} \mat{M} = \mat{M}^\trans \mat{\Sigma}$ according to the definition of $\mat{M}$ in \eqref{expression:M} and the definition of the pseudoinverse of a matrix.}

Since the coefficient matrix of $\mat{\beta}^p$, i.e., $\mat{M}^\trans \mat{\Sigma} \mat{M} \cdot \left({\mat{x}^p}^\trans \left[ \mat{I}_n - \mat{X}_1 \left(\mat{X}_1^\trans \mat{X}_1\right)^{-1} \mat{X}_1^\trans \right] \mat{x}^p \right)$ is positive semidefinite, \eqref{equ:mlr_model_objective_function_simplfied_matrix_form_derivative_equation_2_solution} implies that when
\begin{equation}\label{equ:mlr_model_upper_bound_of_alpha}
\begin{split}
     \alpha &\geq \normof{\mat{M}^\trans \mat{\Sigma} \mat{\widetilde{Y}} \left[ \mat{I}_n - \mat{X}_1 \left(\mat{X}_1^\trans \mat{X}_1\right)^{-1} \mat{X}_1^\trans \right] \mat{x}^p}{\infty} \\
     &= \normof{\mat{\Sigma} \mat{\widetilde{Y}} \left[ \mat{I}_n - \mat{X}_1 \left(\mat{X}_1^\trans \mat{X}_1\right)^{-1} \mat{X}_1^\trans \right] \mat{x}^p}{\infty} \\
     &= \max_j \abs{ \frac{1}{\sigma_i^2} {\mat{\tilde{y}}_i}^\trans \left[ \mat{I}_n - \mat{X}_1 \left(\mat{X}_1^\trans \mat{X}_1\right)^{-1} \mat{X}_1^\trans \right] \mat{x}^p},
\end{split}
\end{equation}
where the next to last equality is due to \eqref{equ:M_trans_Y_tilde_simplfied}, we obtain zero solution.


\subsubsection{Type II Penalty}
The derivative of $p\left(\mat{B} \right) = \alpha \displaystyle{\sum_{i=1}^m} \, \twonorm{\mat{\beta}_i}$ with respect to $\mat{B}$ is
\begin{equation}\label{equ:mlr_model_objective_function_simplfied_matrix_form_derivative_TypeIIPenalty}
\frac{\partial p\left(\mat{B} \right)}{\partial \mat{B}} = \alpha \left(
                                                                    \begin{array}{c}
                                                                      \frac{\partial \twonorm{\mat{\beta}_1}}{\partial \mat{\beta}_1^\trans} \\
                                                                      \frac{\partial \twonorm{\mat{\beta}_2}}{\partial \mat{\beta}_2^\trans} \\
                                                                      \vdots \\
                                                                      \frac{\partial \twonorm{\mat{\beta}_m}}{\partial \mat{\beta}_m^\trans} \\
                                                                    \end{array}
                                                                  \right),
\end{equation}
where $\frac{\partial \twonorm{\mat{\beta}_i}}{\partial \mat{\beta}_i} = \frac{\mat{\beta}_i}{\twonorm{\mat{\beta}_i}}$ if $\mat{\beta}_i \neq \mat{0}$ and $\twonorm{\frac{\partial \twonorm{\mat{\beta}_i}}{\partial \mat{\beta}_i}} \leq 1$ otherwise~\cite{Yuan2006,Friedman2010}.

Setting \eqref{equ:mlr_model_objective_function_simplfied_matrix_form_derivative} equal to zero yields
\begin{eqnarray}
 \mat{X}^\trans\mat{X} \mat{B}^\trans \mat{M}^\trans \mat{\Sigma} \mat{m}^i - \mat{X}^\trans \mat{\widetilde{Y}}^\trans \mat{\Sigma} \mat{m}^i + \alpha \frac{\partial \twonorm{\mat{\beta}_i}}{\partial \mat{\beta}_i} = \mat{0}_{p \times 1},
\end{eqnarray}
for $i=1,2,\dots,m$, where $\mat{m}^i$ is the $i$-th column of $\mat{M}$ in \eqref{expression:M}. The minimizer to $f\left(\mat{B}\right)$ is a zero matrix when
\begin{equation}\label{equ:mlr_model_upper_bound_of_alpha_simplfied_TypeIIPenalty}
     \alpha \geq \max_j \twonorm{\mat{X}^\trans \mat{\widetilde{Y}}^\trans \mat{\Sigma} \mat{m}^i}.
\end{equation}

Note that
\begin{equation}\label{equ:equ:Y_tilde_trans_mi_simplfied}
     \mat{\widetilde{Y}}^\trans \mat{\Sigma} \mat{m}^i = \mat{\widetilde{Y}}^\trans \mat{\Sigma} \left( \mat{e}_i - \frac{1}{\displaystyle{\sum_{\ell=1}^m} \frac{1}{\sigma_\ell^2}} \frac{1}{\sigma_i^2} \mat{1}_m\right) = \mat{\widetilde{Y}}^\trans \mat{\Sigma} \mat{e}_i = \frac{1}{\sigma_i^2}\mat{\tilde{y}}_i,
\end{equation}
where the next to last equality is due to \eqref{equ:y_tilde_sum_over_i}.

Substituting \eqref{equ:equ:Y_tilde_trans_mi_simplfied} into \eqref{equ:mlr_model_upper_bound_of_alpha_simplfied_TypeIIPenalty} yields
\begin{equation}\label{equ:mlr_model_upper_bound_of_alpha_expression_simplified_TypeIIPenalty}
     \alpha_{\rm max} = \max_j \twonorm{ \frac{1}{\sigma_i^2}\mat{X}^\trans\mat{\tilde{y}}_i }.
\end{equation}

\subsection{Estimation of $\{\sigma_i^2\}_{i=1}^m$}
\label{subsec:multiple_linear_regression_model_estimate_noise_variance}

To solve for $\{\sigma_i^2\}_{i=1}^m$, consider the negative log-likelihood function with $\{\sigma_i^2\}_{i=1}^m$ being unknown parameters as well:
\begin{equation}\label{equ:mlr_model_negative_log-likelihood}
l\left(\mat{\beta}_0, \{\mat{\beta}_i\}, \mat{d}; \mat{y}\right) = \displaystyle{\sum_{i=1}^m} \left[ \frac{n}{2} \log(2\pi\sigma_i^2) + \frac{1}{2\sigma_i^2}
 \displaystyle{\sum_{j=1}^n} \left(y_{ij}-\beta_{0i}-\mat{x}_j^\trans \mat{\beta}_i-d_j\right)^2 \right].
\end{equation}

Taking partial derivatives of $l(\cdot)$ with respect to $d_j$ and $\beta_{0i}$ and setting the result to zero, we arrive at 
\begin{equation}\label{equ:mlr_model_d_j}
    d_j = d_j -d_1 = \left(\bar{y}_{\cdot j}^{(w)}-\bar{y}_{\cdot 1}^{(w)}\right) - \left(\mat{x}_j-\mat{x}_1\right)^\trans \mat{\bar{\beta}}^{(w)},
\end{equation}
\begin{equation}\label{equ:mlr_model_mu_i_no_d_j_simplfied}
    \beta_{0i} = \bar{y}_{i\cdot} - \frac{1}{n} \displaystyle{\sum_{j=1}^n} \mat{x}_j^\trans \mat{\beta}_i - \frac{1}{n} \displaystyle{\sum_{j=1}^n} d_j = \bar{y}_{i\cdot} + \bar{y}_{\cdot 1}^{(w)} - \bar{y}^{(w)} - \mat{x}_1^\trans \mat{\bar{\beta}}^{(w)}.
\end{equation}
\rev{where to derive the second equality we have exploited assumption \eqref{equ:mlr_model_assumption}.}

The sum of \eqref{equ:mlr_model_d_j} and \eqref{equ:mlr_model_mu_i_no_d_j_simplfied} gives
\begin{equation}\label{equ:mlr_model_mu_i_plus_d_j}
    \beta_{0i} + d_j = \bar{y}_{i\cdot} + \bar{y}_{\cdot j}^{(w)} - \bar{y}^{(w)} - \mat{x}_j^\trans \mat{\bar{\beta}}^{(w)}.
\end{equation}

Taking partial derivatives of $l(\cdot)$ with respect to $\mat{\beta}_i$ and setting the result to zero, we have
\begin{equation}\label{equ:mlr_model_beta_i_esti}
    \mat{\beta}_i = \displaystyle{\sum_{j=1}^n} \mat{x}_j y_{ij} - \displaystyle{\sum_{j=1}^n} \mat{x}_j \left( \beta_{0i}+d_j \right).
\end{equation}
Substituting \eqref{equ:mlr_model_mu_i_plus_d_j} into \eqref{equ:mlr_model_beta_i_esti} yields
\begin{equation}\label{equ:mlr_model_beta_i_esti_original}
    \mat{\beta}_i = \left(\mat{X}^\trans \mat{X}\right)^{-1} \left[\displaystyle{\sum_{j=1}^n} \mat{x}_j y_{ij} - \frac{1}{\displaystyle{\sum_{i=1}^m} \frac{1}{\sigma_i^2}} \displaystyle{\sum_{i=1}^m} \frac{1}{\sigma_i^2} \displaystyle{\sum_{j=1}^n} \mat{x}_j y_{ij}\right] + \mat{\bar{\beta}}^{(w)},
\end{equation}
where $\mat{\bar{\beta}}^{(w)}$ is defined in \eqref{equ:mlr_model_weighted_average_beta_over_i}.

Taking partial derivatives of $\sigma_i^2$ and setting the result to zero gives
\begin{equation}\label{equ:mlr_model_sigma_i_squared}
    \sigma_i^2 = \frac{1}{n} \displaystyle{\sum_{j=1}^n} \left(y_{ij}-\beta_{0i}-\mat{x}_j^\trans \mat{\beta}_i-d_j\right)^2.
\end{equation}
Substituting \eqref{equ:mlr_model_mu_i_plus_d_j} into \eqref{equ:mlr_model_sigma_i_squared} yields
\begin{equation}\label{equ:mlr_model_sigma_i_squared_original}
    \sigma_i^2 = \frac{1}{n} \displaystyle{\sum_{j=1}^n} \left(y_{ij} - \bar{y}_{i\cdot} - \bar{y}_{\cdot j}^{(w)} + \bar{y}^{(w)} - \mat{x}_j^\trans \mat{\beta}_i + \mat{x}_j^\trans \mat{\bar{\beta}}^{(w)} \right)^2,
\end{equation}
where $\bar{y}_{i\cdot}$, $\bar{y}_{\cdot j}^{(w)}$ and $\bar{y}^{(w)}$ are defined in \eqref{equ:average_y_over_j}, \eqref{equ:weighted_average_y_over_i} and \eqref{equ:weighted_average_y_over_ij}, respectively.

Given initial estimates for $\mat{\bar{\beta}}^{(w)}$ and $\{\sigma_i^2\}_{i=1}^m$, estimates for $\mat{\beta}_i$ and $\sigma_i^2$ can then be iteratively updated using equations \eqref{equ:mlr_model_beta_i_esti_original}, \eqref{equ:mlr_model_sigma_i_squared_original}, and \eqref{equ:mlr_model_weighted_average_beta_over_i} until convergence.

After estimating $\sigma_i^2$'s, they can then be shrinked (squeezed) toward the common noise variance to obtain robust estimates for $\sigma_i^2$, as done in Section \ref{subsec:estimate_noise_variance}.

\section{Experiments}
\label{sec:experiments}

We evaluate the performance of the proposed algorithms. To save space, we only verify the proposed algorithm for the simple regression model \eqref{sec:model} (referred to as slr-ADMM). 
We set the penalty parameter as $\alpha=0.01 \alpha_{\rm max}$, where $\alpha_{\rm max}$ is determined according to~\eqref{equ:simple_regression_model_upper_bound_of_alpha_simplified}.

\subsection{Simple Test with Synthetic Data}

We simulate RNA-seq data with a total of $m=1000$ genes and $n=15$ samples. Other parameters are set as below. 
\begin{table}[ht]
\caption{Nonlinear Model Results} 
\centering 
\begin{tabular}{c c c c} 
\hline\hline 
parameter & meaning \\ [0.5ex] 
\hline 
    $\beta_{0i} \sim \mathcal{N}(-3,2)$ &\mbox{ $y$-intercept}\\
	$\beta_i = 0$ &\mbox{ log-fold change for non-DE genes}\\
	$\beta_i \sim \mathcal{N}(2,1)$ &\mbox{ log-fold change for up-regulated DE genes}\\
    $\beta_i \sim \mathcal{N}(-2,1)$ &\mbox{ log-fold change for down-regulated DE genes}\\
    $x_j \sim \mathcal{N}(0,1)$ &\mbox{ condition data}\\
    $y_{ij}\sim \mathcal{N}(\beta_{0i}+\beta_i x_j, 0.1)$ &\mbox{ log gene expression}\\
	$\log(l_i)\sim {\rm unif}(5,10)$ &\mbox{ log gene lengths} \\
	$N_j \sim Unif(3, 5)\times 10^7$ &\mbox{ library sizes} \\
	$c_{ij} = \left\lfloor N_j \cdot \frac{{l_ie^{y_{ij}}}_{i=1}^m}{\sum_{i=1}^ml_i e^{y_{ij}}}\right\rfloor+1$ &\mbox{ read counts} \\ [1ex] 
\hline 
\end{tabular}
\label{table:nonlin} 
\end{table}

\begin{figure}
\centering
\includegraphics[angle=-90,width=0.40\textwidth]{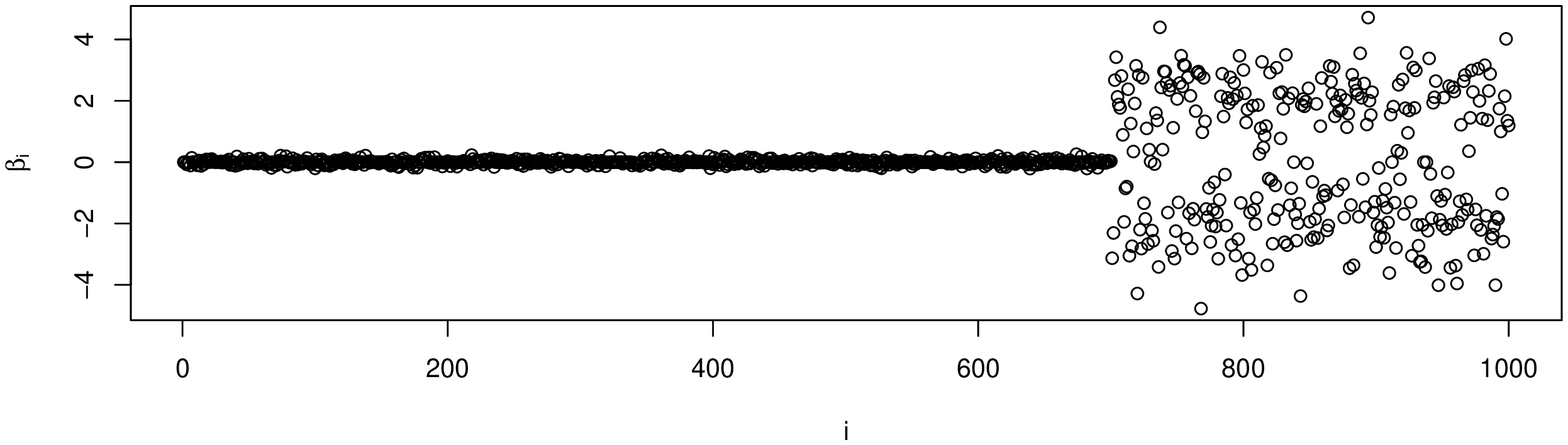}\\
\footnotesize{(a) 300 DE genes, among which 50\% are up-regulated}\\
\includegraphics[angle=-90,width=0.40\textwidth]{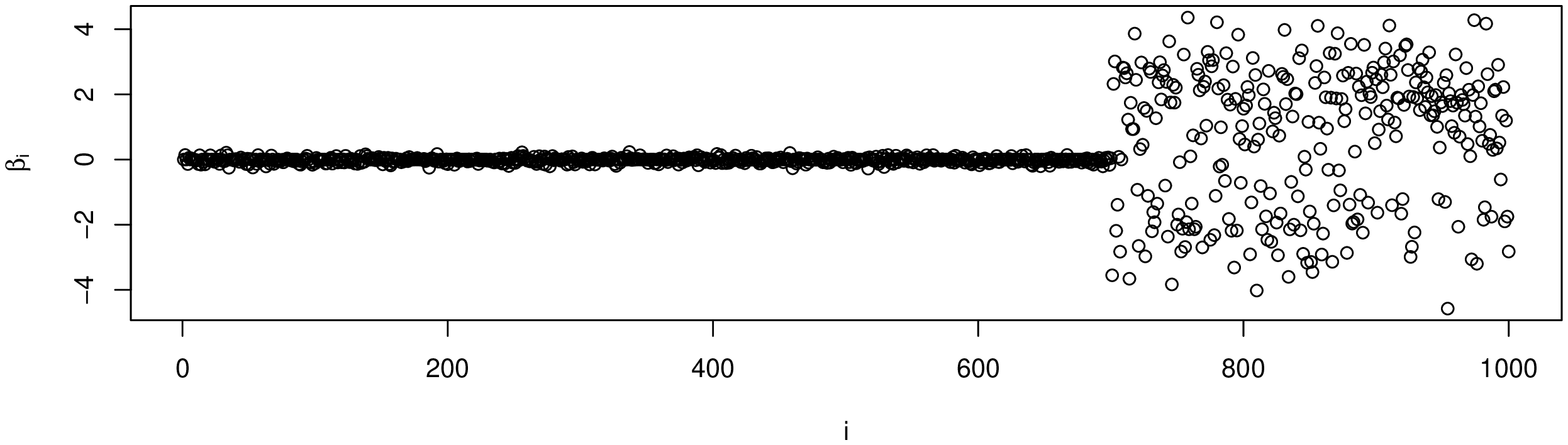}\\
\footnotesize{(b) 300 DE genes, among which 70\% are up-regulated}\\
\includegraphics[angle=-90,width=0.40\textwidth]{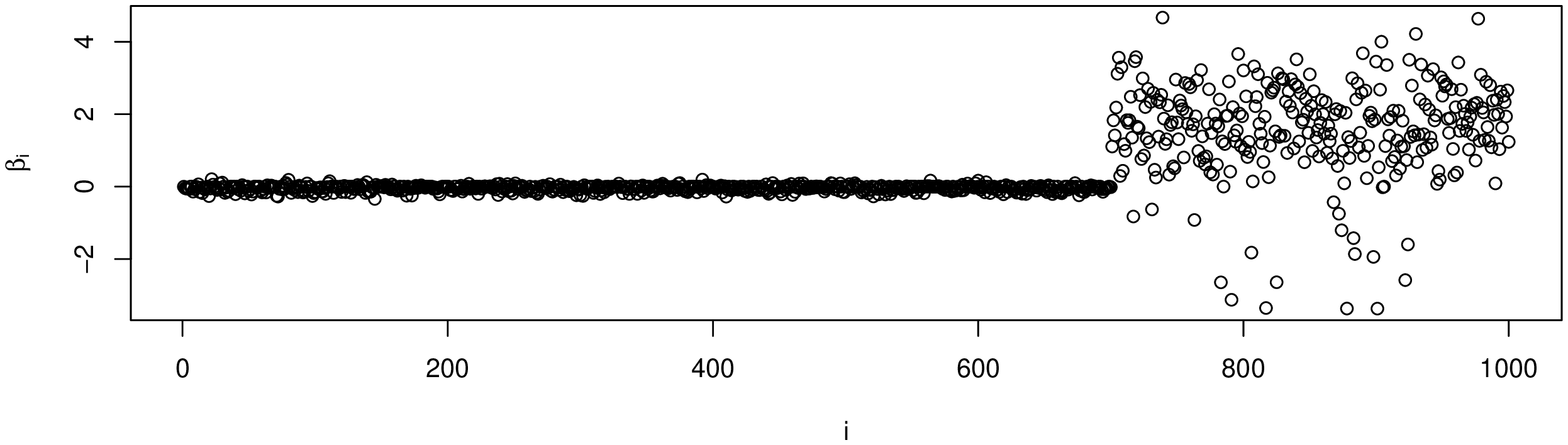}\\
\footnotesize{(c) 300 DE genes, among which 90\% are up-regulated}\\
\includegraphics[angle=-90,width=0.40\textwidth]{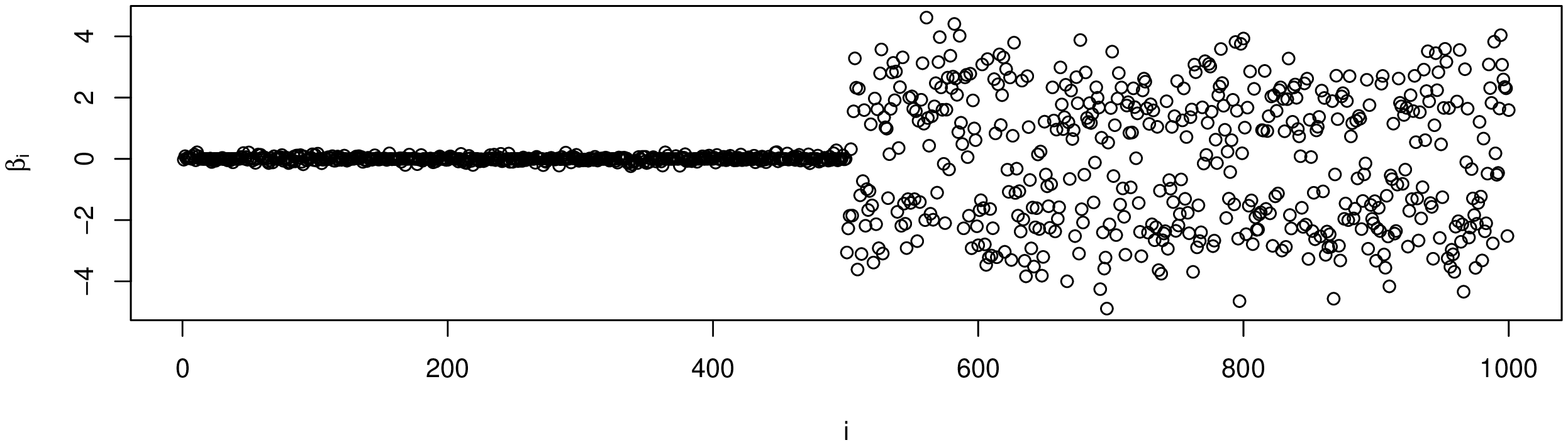}\\
\footnotesize{(d) 500 DE genes, among which 50\% are up-regulated}\\
\includegraphics[angle=-90,width=0.40\textwidth]{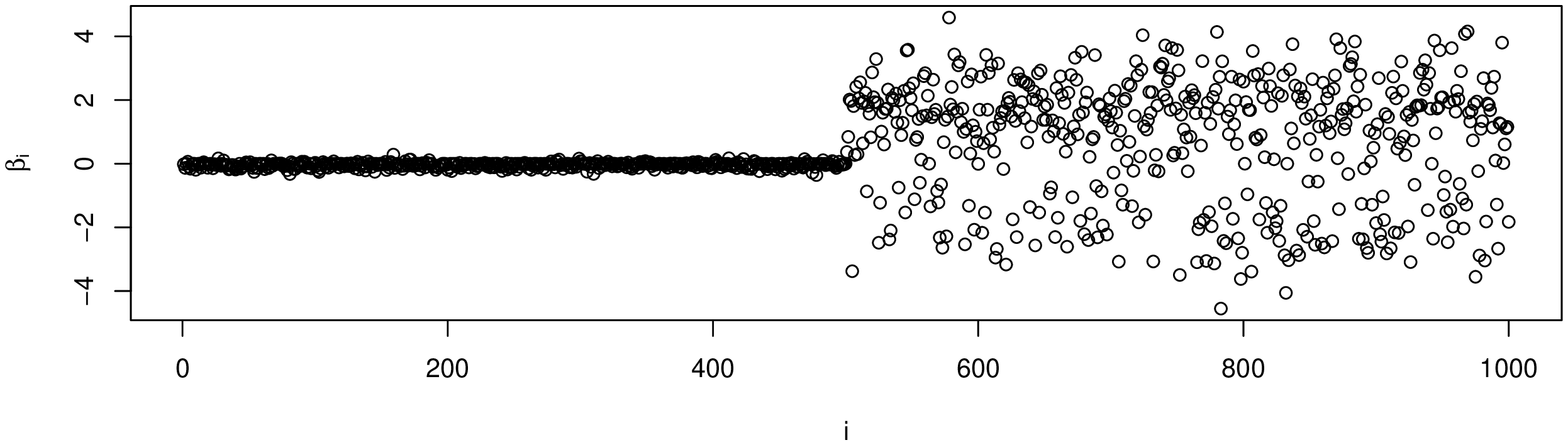}\\
\footnotesize{(e) 500 DE genes, among which 70\% are up-regulated}\\
\includegraphics[angle=-90,width=0.40\textwidth]{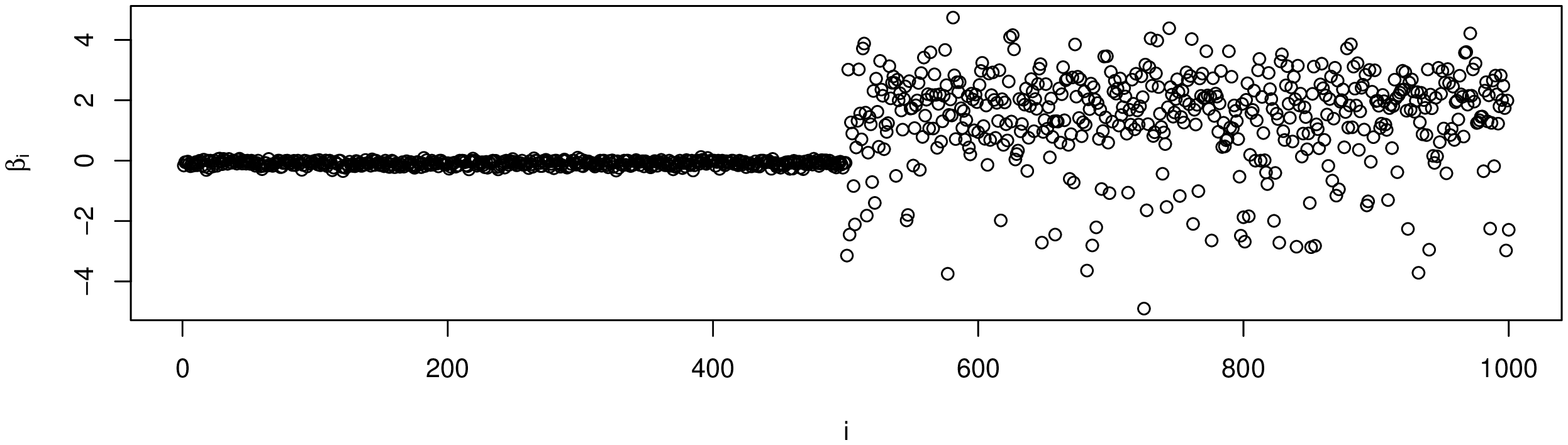}\\
\footnotesize{(f) 500 DE genes, among which 90\% are up-regulated}\\
\includegraphics[angle=-90,width=0.40\textwidth]{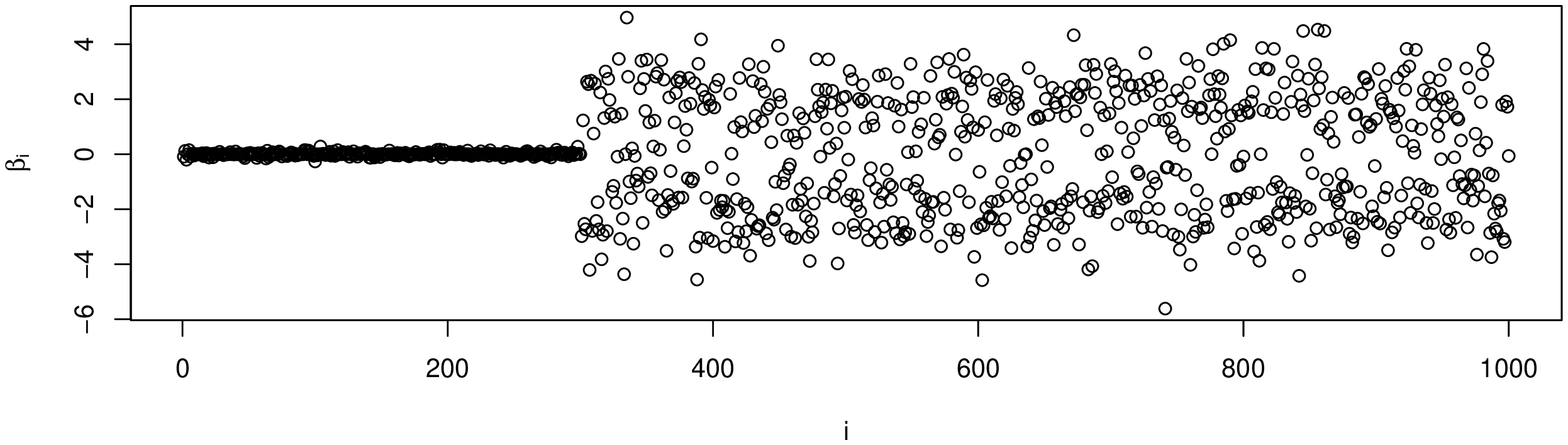}\\
\footnotesize{(g) 700 DE genes, among which 50\% are up-regulated}\\
\includegraphics[angle=-90,width=0.40\textwidth]{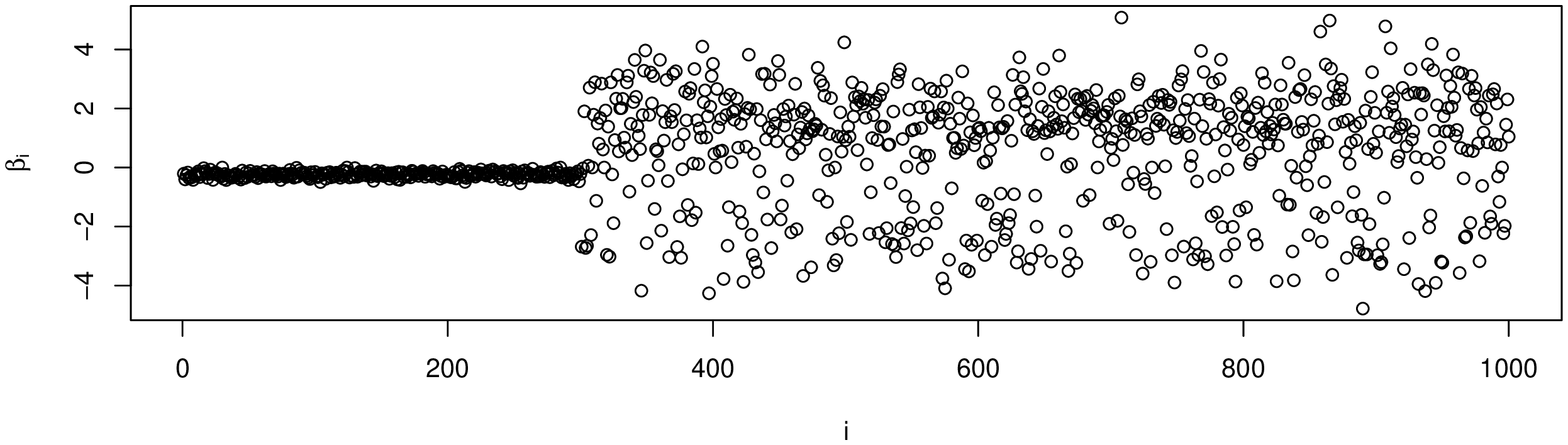}\\
\footnotesize{(h) 700 DE genes, among which 70\% are up-regulated}\\
\includegraphics[angle=-90,width=0.40\textwidth]{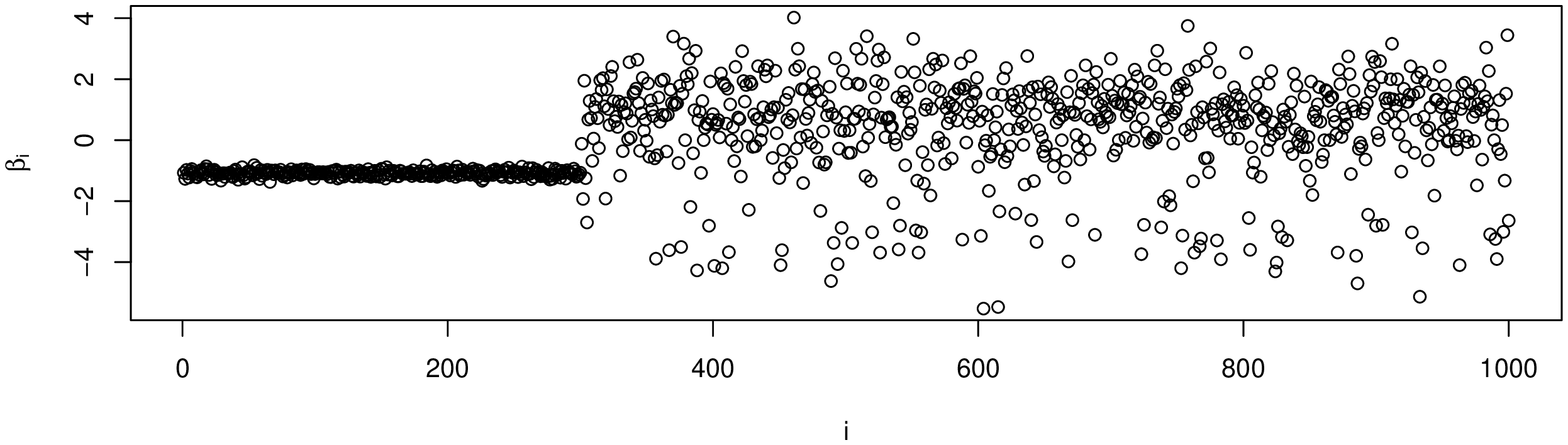}\\
\footnotesize{(i) 700 DE genes, among which 90\% are up-regulated}\\
\caption{Estimated $\beta_i$ from simulated data with simple linear regression model. The number of genes is $m=1000$ and number of samples is $n=15$. The penalty parameter is set as $\alpha=0.01 \alpha_{\rm max}$, where $\alpha_{\rm max}$ is determined according to~\eqref{equ:simple_regression_model_upper_bound_of_alpha_simplified}.}
\label{fig:slr_m=1000_n=15}
\end{figure}

We first simulate with 700 non-DE genes and 300 DE genes. Among DE genes 50\% are up-regulated while the remaining 50\% are down-regulated. The fitted $\{\beta_i\}_{i=1}^m$ using slr-ADMM are plotted in Figure~\ref{fig:slr_m=1000_n=15}(a). We see that the non-DE genes corresponds to exactly zero or close-to-zero coefficients while DE genes corresponds to large non-zero coefficients and they are easily distinguishable from each other. In Figure~\ref{fig:slr_m=1000_n=15}(b), we increase the percent of up-regulated DE genes to 70\%. Our method still retrieves almost all non-zero $\beta_i$'s while shrinking all other $\beta_i$'s to zero. We further increase the percent of up-regulated DE genes to 90\%, for which our method still achieves accurate estimates [Figure~\ref{fig:slr_m=1000_n=15}(c)].

In Figure~\ref{fig:slr_m=1000_n=15}(d-f), we increase the number of DE genes to 500, among which 50\%, 70\% or 90\% are up-regulated while others are down-regulated. Our method still achieves accurate estimates. In Figure~\ref{fig:slr_m=1000_n=15}(g-h), we further increase the number of DE genes to 700 among which 50\% or 70\% are up-regulated, for which our method still achieves accurate estimates when. Only when we simulate with 700 DE genes among which 90\% are up-regulated, our method fails to distinguish between DE and non-DE genes since the estimated regression coefficients of the latter are not zero either [Figure~\ref{fig:slr_m=1000_n=15}(i)]

Using a different gene expression measure such as CPM, RPKM or TPM values computed with formulas in~(\ref{units}) yields essentially the same result.

\subsection{Comparison with Existing Methods with Real Data}
We compare our method with edgeR-robust~\cite{edgeR,edgeR-robust}, DESeq2~\cite{DESeq2}, and voom~\cite{limma,limma-voom}, all of which are state-of-the-art methods for detecting differential gene expression from RNA-seq data.

We use RNA-seq data with a total of $m = 1000$ genes and $n=15$ samples. We simulate both log-normally distributed read counts, which is the model assumptions of voom and slr-ADMM, as well as negative-binomially distributed read counts, which is the underlying assumption of edgeR-robust and DESeq2. The gene expression levels and library sizes for both simulations are generated based on a real RNA-seq dataset~\cite{Pickrell2010}. The read counts are simulated according to\rev{[~\cite{edgeR-robust}]}. We slightly modify the simulator to allow log-normally distributed data, as well as variable fold changes. We simulate data sets with 30\%, 50\%, 70\%, or 90\% DE genes, among which 50\%, 70\% or 90\% are up-regulated while others are down-regulated. The log-fold changes for unit changes of explanatory variables, i.e., $\beta_i$'s, for up-regulated DE genes are assumed to be distributed as $\mathcal{N}(1, 1)$, while that for down-regulated DE genes are distributed as $\mathcal{N}(-1, 1)$. \rev{We consider high signal-to-noise ratio (low noise level) scenarios~\footnote[3]{\rev{For low SNRs, the simulation results are essentially similar except that the relative performance gain or loss becomes less significant and non-typical, which renders it difficult to interpret.}}}, and set the negative-binomial dispersion parameter to one fifth of that generated from the real RNA-seq dataset. For log-normal data simulation, log read counts are assumed to be normally distributed with $\sigma=0.1$.

The AUCs for DE gene detection from log-normally and negative-binomially distributed data using all four methods are summarized in Tables~\ref{tab:n=15_LN} and ~\ref{tab:n=15_NB}, respectively.


In Table~\ref{tab:n=15_LN}, we see that the voom and slr-ADMM perform the best with log-normally distributed data. In relatively easier cases where only a small proportion of genes are differentially expressed, in a symmetric manner (which means the numbers of up- and down-regulated genes among DE genes are roughly the same), the voom and DESeq2 perform the best. The slr-ADMM is slightly inferior to the best performer but by only a small margin (i.e., within approximately one standard error for the log-normal distributed datasets). In challenging cases where a large proportion of genes are differentially expressed in a asymmetric manner (e.g., when more than 75\% genes are up-regulated), the slr-ADMM performs the best. And with the increase of the percent of DE genes and/or percent of up-regulated genes, the performance gain of the slr-ADMM over completing methods increases and becomes more significant. 

For negative-binomially distributed data, in Table~\ref{tab:n=15_NB} we see that the edgeR-robust and DESeq2 perform the best in easy cases. But again, in challenging scenarios where a large proportion of genes are differentially expressed in a asymmetric manner, the slr-ADMM is superior to all other methods.


\begin{table}
\caption{AUC comparison of edgeR-robust, DESeq2, limma and slr-ADMM in log-normally distributed data. The number of samples is $n=15$. The table shows the percent of DE genes (DE \%), percent of up-regulated genes among all the DE genes (Up \%), as well as the mean AUCs for all four methods measured using 10 simulated replicates. The standard errors of the mean AUCs are given in parentheses. \label{tab:n=15_LN}}
\centering
\begin{tabular}{cccccc}
  \hline
DE (\%) & Up (\%) & edgeR-robust & DESeq2 & voom & slr-ADMM \\
  \hline
30 & 50 & 0.954 & 0.955 & {\textbf{0.962}} & 0.9604 \\
    &   & (0.0023) & (0.0021) & {\textbf{(0.002)}} & (0.0016) \\
  30 & 70 & 0.9476 & 0.9506 & 0.9585 & {\textbf{0.9615}} \\
    &   & (0.004) & (0.0039) & (0.004) & {\textbf{(0.004)}} \\
  30 & 90 & 0.9298 & 0.9365 & 0.9404 & {\textbf{0.9554}} \\
    &   & (0.0054) & (0.0029) & (0.0043) & {\textbf{(0.0018)}} \\
  50 & 50 & 0.9498 & 0.9516 & {\textbf{0.9607}} & 0.9593 \\
    &   & (0.0023) & (0.0022) & {\textbf{(0.0023)}} & (0.0022) \\
  50 & 70 & 0.9214 & 0.9333 & 0.9357 & {\textbf{0.9558}} \\
    &   & (0.0031) & (0.0024) & (0.0029) & {\textbf{(0.0029)}} \\
  50 & 90 & 0.8661 & 0.8852 & 0.8926 & {\textbf{0.9499}} \\
    &   & (0.0066) & (0.0045) & (0.0055) & {\textbf{(0.0013)}} \\
  70 & 50 & 0.9482 & 0.9498 & {\textbf{0.9574}} & 0.9564 \\
    &   & (0.0023) & (0.0023) & {\textbf{(0.0019)}} & (0.0023) \\
  70 & 70 & 0.8556 & 0.8925 & 0.8818 & {\textbf{0.9467}} \\
    &   & (0.0031) & (0.0036) & (0.0033) & {\textbf{(0.0023)}} \\
  70 & 90 & 0.6936 & 0.7598 & 0.7223 & {\textbf{0.8587}} \\
    &   & (0.0083) & (0.0074) & (0.0074) & {\textbf{(0.0074)}} \\
   \hline
\end{tabular}
\end{table}

\begin{table}
\caption{AUC comparison of edgeR-robust, DESeq2, limma and slr-ADMM in negative-binomially distributed data. See Table~\ref{tab:n=15_LN} for descriptions.} \label{tab:n=15_NB}
\centering
\begin{tabular}{cccccc}
  \hline
DE (\%) & Up (\%) & edgeR-robust & DESeq2 & voom & slr-ADMM \\
  \hline
30 & 50 & 0.8935 & {\textbf{0.8942}} & 0.8941 & 0.8909 \\
    &   & (0.0051) & {\textbf{(0.0046)}} & (0.0042) & (0.0042) \\
  30 & 70 & {\textbf{0.892}} & 0.8914 & 0.8897 & 0.8915 \\
    &   & {\textbf{(0.0038)}} & (0.0046) & (0.0043) & (0.0045) \\
  30 & 90 & 0.8711 & 0.8707 & 0.868 & {\textbf{0.8875}} \\
    &   & (0.0047) & (0.0034) & (0.0046) & {\textbf{(0.0038)}} \\
  50 & 50 & 0.9116 & {\textbf{0.9119}} & 0.9109 & 0.9074 \\
    &   & (0.0024) & {\textbf{(0.0027)}} & (0.0024) & (0.0025) \\
  50 & 70 & 0.8741 & 0.8751 & 0.8702 & {\textbf{0.8872}} \\
    &   & (0.0043) & (0.0032) & (0.0043) & {\textbf{(0.0034)}} \\
  50 & 90 & 0.8096 & 0.8166 & 0.8043 & {\textbf{0.8695}} \\
    &   & (0.0052) & (0.0035) & (0.006) & {\textbf{(0.0034)}} \\
  70 & 50 & {\textbf{0.9039}} & 0.9024 & 0.9024 & 0.8992 \\
    &   & {\textbf{(0.0023)}} & (0.0019) & (0.002) & (0.0018) \\
  70 & 70 & 0.8493 & 0.8423 & 0.8421 & {\textbf{0.8742}} \\
    &   & (0.0044) & (0.0054) & (0.0047) & {\textbf{(0.0028)}} \\
  70 & 90 & 0.661 & 0.661 & 0.6535 & {\textbf{0.7375}} \\
    &   & (0.0079) & (0.0073) & (0.0073) & {\textbf{(0.0067)}} \\
   \hline
\end{tabular}
\end{table}

Note that when more samples are available (e.g., $n=25$), the performance gain of the slr-ADMM over completing methods becomes even more significant, for both log-normally and negative-binomially distributed data. \rev{For sake of conciseness, the results are not shown here.}

\section{Discussion}
\label{sec:discussion}
 A unified statistical model is proposed for joint between-sample normalization and DE detection of RNA-seq data. The sample-specific normalization factors are modeled as unknown parameters and jointly estimated together with DE detection. As a result, the model is robust against normalization errors and is independent of the units (i.e., counts, CPM/RPM, RPKM/FPKM or TPM) in which gene expression levels are summarized.

For the model with a single treatment condition, we introduce the $\ell_1$-norm penalty to the linear regression model. The $\ell_1$-norm penalty favors sparse solutions (forces some coefficients to be exactly zero). This is desirable since many genes are not differentially expressed. From a Bayesian point of view, the lasso penalty corresponds to a Laplace (double exponential, with zero-mean) prior over the regression coefficients. \rev{By contrast, existing methods do not exploit the sparsity-inducing prior information. 
For the model with multiple treatment conditions, two types of penalty functions are introduced. In the first one only one covariate is of interest while all other covariates are treated as confounding factors. We are interested in testing whether that specific covariate is associated with differential expression. In the second case all covariates are of interest (there are no confounding covariates) and we are interested in testing whether any covariate affects the differential expression of a gene. Regarding choice of the penalty parameter, we theoretically derive the maximum penalty parameter $\alpha_{\rm max}$ that leads to all-zero solution, and set $\alpha=\epsilon \alpha_{\rm max}$ with $0<\epsilon<1$. Empirically we found that the performance is not sensitive to $\epsilon$ and setting $\epsilon=0.01$ works quite well in a wide range of parameter settings. \rev{This avoids computationally expensive cross validation procedure to tune the penalty parameter.}}

Simulation studies show that the proposed methods perform better than or comparably to existing methods in terms of AUC. The performance gain is more significant when a large proportion of genes (e.g., more than half) are differentially expressed and/or the up- and down-regulated DE genes are unbalanced in number, particularly in the presence of high signal-to-noise ratios or large sample-size. 


The R as well as MATLAB codes of the algorithms described in the paper are available for download at \url{http://www-personal.umich.edu/~jianghui/lr-ADMM/}.


%

%


The authors would like to thank...

\ifCLASSOPTIONcaptionsoff
  \newpage
\fi



\bibliographystyle{IEEEtran}
\bibliography{IEEEabrv,RNASeqDEA}

\end{document}

%% file: macros.tex
\usepackage{MnSymbol}
\newcommand{\mat}[1]{\mbox{\boldmath$#1$}} 

\newcommand{\real}{\mathbb{R}}         

\newcommand{\vecelem}[2]{\errmessage{DO NOT USE}}    

\newcommand{\normof}[2]{\left\|#1\right\|_{#2}}
\newcommand{\fronorm}[1]{\normof{#1}{\rm F}}            

\newcommand{\pinv}{\dagger}          
\newcommand{\trans}{{\rm T}}   

\usepackage{color}
